\documentclass{article}

\usepackage{arxiv}

\usepackage{amsmath, amssymb, bm, epsfig}
\usepackage{bm}
\usepackage{bbm}
\usepackage{graphicx}
\usepackage{srcltx}
\usepackage{rotating}
\usepackage{lscape}
\usepackage{booktabs, subfigure}
\usepackage{verbatim, setspace}
\usepackage{calc, color, setspace}
\usepackage[english]{babel}
\usepackage{tabularx}
\usepackage{longtable}
\usepackage{multirow}
\usepackage{booktabs}
\usepackage{ragged2e}
\usepackage{subfigure}
\usepackage{tikz}
\usepackage{ragged2e}
\usepackage{times}
\usepackage{natbib}

\usepackage{url}
\usepackage{rotating}

\newcommand{\CG}[1]{\textcolor{black}{#1}}

\newcommand{\ind}{\stackrel {{\rm ind.}}{\sim}}

\newcommand{\sumas}{\sum^n_{i=1}}

\newcommand{\ii}{i\in\{1,\ldots,n\}}

\newtheorem{proposition}{Proposition}

\newcommand{\se}{\mathbf{s}}

\newcommand{\bI}{\mathbf{I}}
\newcommand{\bmu}{\mbox{\boldmath $\mu$}}

\newcommand{\bnu}{\mbox{\boldmath $\nu$}}

\newcommand{\bsigma}{\mbox{\boldmath $\sigma$}}
\newcommand{\bSigma}{\mbox{\boldmath $\Sigma$}}

\newcommand{\bbeta}{\mbox{\boldmath $\beta$}}
\newcommand{\btheta}{\mbox{\boldmath $\theta$}}

\newcommand{\x}{\mathbf{x}}
\newcommand{\w}{\mathbf{w}}

\newcommand{\A}{\mathbf{A}}
\newcommand{\B}{\mathbf{B}}
\newcommand{\D}{\mathbf{D}}
\newcommand{\bV}{\mathbf{V}}

\newcommand{\bgamma}{\mbox{\boldmath $\gamma$}}
\newcommand{\bGamma}{\mbox{\boldmath $\Gamma$}}

\newcommand{\be}{\mathbf{b}}

\newcommand{\bC}{\mathbf{C}}
\newcommand{\yp}{\mathbf{y}}
\newcommand{\xp}{\mathbf{x}}
\newcommand{\y}{\mathbf{y}}
\newcommand{\Y}{\mathbf{Y}}

\newcommand{\C}{\mathbf{C}}

\newcommand{\Z}{\mathbf{Z}}

\newcommand{\X}{\mathbf{X}}
\newcommand{\bW}{\mathbf{W}}
\newcommand{\bw}{\mathbf{w}}

\newcommand{\E}{\textrm{E}}
\newcommand{\ap}{\mathbf{a}}
\newcommand{\bp}{\mathbf{b}}
\newcommand{\bx}{\mathbf{x}}
\newcommand{\bu}{\mathbf{u}}
\newcommand{\tr}{\textrm{tr}}
\newcommand{\binfty}{\boldsymbol{\infty}}
\newcommand{\RR}{\mathbbm{R}}

\newcommand{\EE}{\mathbbm{E}}
\newcommand{\W}{\mathbf{W}}
\newcommand{\T}{\textrm{T}}

\newcommand{\CB}[1]{\textcolor{black}{#1}}

\title{Heckman selection-t model: parameter estimation via the EM-algorithm}

\author{
 Victor H. Lachos Davila \\
  Department of Statistics\\
  University of Connecticut\\
  Storrs, CT 06269 \\
  \texttt{hlachos@uconn.edu} \\
   \And
 Marcos O. Prates \\
  Departamento de Estad\'istica\\
  Universidad Federal de Minas Gerais\\
  Minas Gerai, MG, Brazil \\
  \texttt{marcosop@est.ufmg.br} \\
  \And
 Dipak K. Dey \\
  Department of Statistics\\
   University of Connecticut\\
  Storrs, CT 06250 \\
  \texttt{dipak.dey@uconn.edu} \\
}

\begin{document}
\maketitle
\begin{abstract}
Heckman selection model is perhaps the most popular econometric model in the analysis of data with sample selection. The analyses of this model are based on the normality assumption for the error terms, however, in some applications, the distribution of the error term departs significantly from normality, for instance, in the presence of heavy tails and/or atypical observation.  In this paper, we explore the Heckman selection-t model where the random errors follow a bivariate Student's-t distribution.  We develop an analytically tractable and efficient EM-type algorithm for iteratively computing maximum likelihood estimates of the parameters, with standard errors as a by-product. The algorithm has closed-form expressions at the E-step, that rely on formulas for the mean and variance of the truncated Student's-t distributions. Simulations studies show the vulnerability of the Heckman selection-normal model, as well as the robustness aspects of the Heckman selection-t model. Two real examples are analyzed, illustrating the usefulness of the proposed methods. The proposed algorithms and methods are implemented in the new \verb"R" package \verb"HeckmanEM".
\end{abstract}

\keywords{ EM-type algorithms \and Heckman selection model \and Multivariate Student's-$t$ \and Robustness.}

\section{Introduction}
Sample selection (SL) often occurs many fields, like economics, biostatistics, finance, medical surveys, sociology and political science, to name a few. For example, in a sample of individual consumers with expenditures below or above some threshold, where the unobserved random variable (decision to spend) is related to the spending amount. As expenditure is not independent of the decision to spend, this sample may represent only a subset of the full population, and thus selection bias arises. In a classic example, \cite{heckman1979} proposed the SL model, aiming to estimate the wage offer function of women. Because housewives’ wages are not observed, the sample collected is subject to the self-selection problem. SL is a special
case of a more general concept known in the econometrics
literature as limited dependent variables or variables observed over a limited range of their support, however, censoring \citep{massuia2015influence} is much simpler than SL.

The classical Heckman SL model was introduced by
\cite{heckman1974shadow} when he proposed a parametric approach to the parameter estimation under the assumption of bivariate normality (SLn). However, it is well-known that several
phenomena are not always in agreement with this assumption, yielding
{residuals} with a distribution with heavy  tails or skewness. These characteristics can be circumvented by data
transformations, as proposed by \cite{lee1983generalized}, which can render
approximate normality with reasonable empirical results. However,
some possible drawbacks of these methods are: (i) transformations
provide reduced information on the underlying data generation
scheme; (ii) component wise transformations may not guarantee joint
normality; (iii) parameters may lose interpretability in a
transformed scale and (iv) transformations may not be universal and
usually vary with the data set. {Hence, from a practical perspective,
	there is a need to seek an appropriate theoretical model that avoids
	data transformations.} 

There are two-ways of estimating the SLn model, via maximum likelihood (ML) and using a two-step procedure \citep{heckman1979}. A drawback of the ML estimation is less robust than the two-step procedure and is sometimes difficult to get it to converge \citep{wooldridge2010econometric}. However, ML estimation will be more efficient if the random error really are jointly normally distributed. In this context, many robust methods have been proposed over the years to broaden the applicability of the SLn model to situations where the Gaussian error term
assumption may be inadequate. For instance, the semiparametric SL model proposed by \cite{ahn1993semiparametric} and the nonparametric SL model proposed by \cite{das2003nonparametric}. From a Bayesian perspective,  \cite{kai1998bayesian} proposed a Bayesian inference procedure for the SL model using data augmentation. More recently, \cite{kim2019bayesian} proposed a flexible nonparametric SL model using Bernstein Polynomial.

In the context of parametric models, \cite{marchenko2012heckman} introduced the Heckman selection-t model (SLt) model that extends the conventional SLn model by \cite{heckman1979} to have a bivariate Student's t error distribution. This model provides a greater flexibility for modeling heavier-tailed data than the SLn model by introducing only one extra parameter, the degrees of freedom, controlling the tails of the distribution. The authors considered  ML estimation of the parameters via Newton-Raphson procedures available in statistical packages such as \verb"R" and \verb"stata". They demonstrated the robustness aspects of the SLt model against outliers through extensive simulations. More recently, \cite{zhao2020new} have proposed EM-type algorithms for ML estimation of the SLn model, which have the advantages of easy implementation and numerical stability. Moreover, ML estimation via the EM algorithm yield better estimators than the 2-step procedure.

Motivated by  \cite{zhao2020new}, in this paper we  propose  a novel, simple and efficient EM-type algorithm for iteratively computing ML estimates of the parameters in the SLt model. We show that the E-step reduces to computing the first two moments of a truncated multivariate Student's t distribution. The general formulas for these moments were derived efficiently by \cite{GalarzaCran}, for which we use the \verb"MomTrunc" package in \verb"R". The likelihood function is easily computed as a byproduct of the E-step and is used for monitoring convergence and for model selection (AIC and BIC). Furthermore, we consider a general information-based method for obtaining the asymptotic covariance matrix of the ML estimate. The method proposed in this paper is implemented in the \verb"R" package \verb"HeckmanEM", which is available for download from Github (\url{https://github.com/marcosop/HeckmanEM}).

The remainder of the paper is organized as follows. In Section~\ref{model}, we
briefly discuss some preliminary results related to the multivariate Student's-$t$ distribution and its truncated
version.  Some of its key properties are also discussed.  In Section~\ref{SLnModel}, we present the SLn model  proposed by \cite{heckman1979} and
the related EM algorithm for ML estimation. In Section~\ref{SLtModel}, we present the robust
SLt model, including the EM-type algorithm for ML estimation, and
derive the empirical information matrix  analytically to obtain the standard errors. In Sections~\ref{secSim} and \ref{secApp}, numerical examples using both
simulated and real data are given to illustrate the performance of the proposed method. Finally, some concluding remarks are presented in Section~\ref{sec:6}.

\section{Background}\label{model}
In this section, we present some useful results associated with the
$p$-variate  Student's-$t$ distribution and its truncated  version,  the $p$-variate  truncated Student's-$t$ distribution. We begin our exposition by defining the notation and presenting some basic concepts which are used throughout the development of our methodology. As is usual in probability theory
and its applications, we denote a random variable by an
\CG{upper-case} letter and its realization by the corresponding
lower case \CG{and} use boldface letters for vectors and
matrices. Let $\mathbf{I}_p$ represent a $p\times p$
identity matrix, $\A^{\top}$ be the transpose of $\A$.  For multiple \CG{integrals, we
	use} the \CG{shorthand} notation
$$\int_{\ap}^{\bp}f(\x)d\x=\int_{a_1}^{b_1}\ldots\int_{a_p}^{b_p}f(x_1,\ldots,x_p)\mathrm{d} x_p\ldots \mathrm{d} x_1.$$
where  $\mathbf{a}=(a_1,\ldots,a_p)^\top$  and $\mathbf{b}=(b_1,\ldots,b_p)^\top$. If the Borel set  in $\mathbb{R}^p$ has the form
\begin{equation} \label{hyper1}
\mathbb{A} = \{(x_1,\ldots,x_p)\in \mathbb{R}^p:\,\,\, a_1\leq x_1 \leq b_1,\ldots, a_p\leq x_p \leq b_p \}=\{\mathbf{x}\in\mathbb{R}^p:\mathbf{a}\leq\mathbf{x}\leq\mathbf{b}\}.
\end{equation}
we  use the notation $\{\Y \in \mathbb{A}\}=\{\ap\leq \Y\leq\bp\}$.
\subsection{The multivariate Student's-$t$  distribution}

A random variable $\X$ having a $p$-variate Student's-$t$ distribution with
location vector $\bmu$, \CG{positive-definite scale-covariance}
matrix $\bSigma$ and degrees of freedom $\nu$, denoted by $\X \sim
t_p(\bmu,\bSigma,\nu)$, has \CG{the pdf:}
$$
t_p(\xp\mid\bmu,\bSigma,\nu)=
\frac{\Gamma(\frac{p+\nu}{2})}{\Gamma(\frac{\nu}{2})\pi^{p/2}}\nu^{-p/2}|\bSigma|^{-1/2}\left(1+\frac{\delta(\xp)}{\nu}\right)^{-(p+\nu)/2},\label{lsdefAB1}
$$
where $\Gamma{(\cdot)}$ is the standard gamma function and $\delta(\xp) = (\xp-\bmu)^{\top}\bSigma^{-1}(\xp-\bmu)$ is the squared Mahalanobis
distance. Let $T_p(\ap,\bp;\bmu,\bSigma,\nu)$ represent
$$T_p(\ap,\bp;\bmu,\bSigma,\nu)=\int_{\ap}^{\bp}{{t}_p}(\x|\bmu,\bSigma,\bnu)\textrm{d}\bx,$$ where  $\mathbf{a}=(a_1,\ldots,a_p)^\top$  and $\mathbf{b}=(b_1,\ldots,b_p)^\top$. When $\ap=-\binfty$ we will write simply $T_p(\bp;\bmu,\bSigma,\nu)$ and when $p=1$ we will omit the sub-index $p$.

It is \CB{known} that  as {$\nu \to \infty$,} $\X$ converges in
distribution to a multivariate normal with mean $\bmu$ and variance-covariance matrix $\bSigma$, denoted by
$N_p(\bmu,\bSigma)$. An important property of the random vector $\X$ is that it can be
written as a scale mixture  of \CG{the MVN} random vector
\CG{coupled with} a positive random variable, {i.e.},
\begin{equation}\label{stoNI1}
\X=\bmu+U^{-1/2}\Z,
\end{equation}
where $\Z \sim N_p(\mathbf{0},\bSigma)$, and is independent of
$U\sim \mbox{Gamma}(\nu/2,\nu/2)$, where $\mbox{Gamma}(a,b)$ denotes a
gamma distribution with mean $a/b$.


The following properties of the  $p$-variate Student's-$t$ distribution are useful for
our theoretical developments. We start with the
marginal-conditional decomposition of a  $p$-variate Student's-$t$ random vector. The proof of the following propositions can be found in
\cite{ArellanoBolfarine95}.

\begin{proposition}\label{prop1}
	Let  $\X\sim t_p(\bmu,\bSigma,\nu)$ \CG{partitioned} as
	$\X^{\top}=(\X^{\top}_1,\X^{\top}_2)^{\top}$ \CG{with} $dim(\X_1) =
	p_1$, $dim(\Y_2) = p_2$, \CG{where} $p_1 + p_2 = p$. Let
	$\bmu=(\bmu^{\top}_1,\bmu^{\top}_2)^{\top}$ and
	$\bSigma=\begin{bmatrix}\bSigma_{11} & \bSigma_{12} \\
	\bSigma_{21} & \bSigma_{22}\end{bmatrix}$ be the corresponding
	partitions of \CG{$\bmu$ and $\bSigma$}. Then, we have
	\begin{itemize}
		\item[$(i)$] $\X_1\sim t_{p_1}(\bmu_1,\bSigma_{11},\nu)$; and
		\item[$(ii)$] The
		conditional distribution of $\X_2 \mid \CG{(\X_1=\xp_1)}$ is given by
		$$
		\X_2 \mid \CG{(\X_1=\xp_1)}\sim t_{p_2}\left(\mathbf{y}_2 \mid\bmu_{2.1},\widetilde{\bSigma}_{22.1},\nu+p_1
		\right),
		$$
		where \CG{$\bmu_{2.1}=\bmu_2+\bSigma_{21}\bSigma^{-1}_{11}(\xp_1-\bmu_1)$
			and}
		$\widetilde{\bSigma}_{22.1}=\left(\displaystyle\frac{\nu+\delta_1}{\nu+p_1}\right)\bSigma_{22.1}$
		\CG{with }$\delta_1=(\xp_1-\bmu_1)^{\top}\bSigma_{11}^{-1}(\xp_1-\bmu_1)$
		\CG{and }
		$\bSigma_{22.1}=\bSigma_{22}-\bSigma_{21}\bSigma^{-1}_{11}\bSigma_{12}$.
	\end{itemize}
\end{proposition}

\begin{proposition}\label{prop11}
	Let  $\X\sim t_p(\bmu,\bSigma,\nu)$. Then for any fixed vector $\be\in \mathbb{R}^ m$ and matrix $\A\in \mathbb{R}^ {m\times p}$ of full rank we get
	$$\bV=\be+\A\X\sim t_m(\be+\A\bmu, \A\bSigma\A^ {\top},\nu).$$
\end{proposition}

\subsection{The multivariate truncated Student's-$t$  distribution}

A $p$-dimensional random vector $\Y$ is said to follow a
doubly truncated Student's-$t$ distribution with location vector
$\bmu$, scale-covariance matrix $\bSigma$ and degrees of freedom
$\nu$ over the truncation region $\mathbb{A}$ defined in
(\ref{hyper1}), denoted by $\mathbf{Y}\sim
Tt_{p}(\bmu,\bSigma,\nu;\mathbb{A})$, if it has the pdf:
$$Tt_p(\yp|\bmu,\bSigma,\nu;\mathbb{A})=\frac{t_p(\yp|\bmu,\bSigma,\nu)}{T_p(\ap,\bp;\bmu,\bSigma,\nu)},\,\,\ap\leq \yp \leq \bp.$$
The cdf of $\Y$ evaluated at the \CG{region} $\ap\leq \yp \leq
\bp$ is
$$TT_p(\yp|\bmu,\bSigma,\nu;\mathbb{A})=\frac{1}{T_p(\ap,\bp;\bmu,\bSigma,\nu)}\int_{\ap}^{\yp}t_p(\xp|\bmu,\bSigma,\nu)d \xp=\frac{T_p(\ap,\yp;\bmu,\bSigma,\nu)}{T_p(\ap,\bp;\bmu,\bSigma,\nu)}.$$

The following propositions are related to the marginal and
conditional moments of the first two moments of the TMVT
distributions under a double truncation. The proof is similar
to those given in  \cite{Matos.SINICA}. \CB{In what follows, we shall use the notation $\Y^{(0)}= 1$, $\Y^{(1)}=\Y$,
	$\Y^{(2)}=\Y\Y^{\top}$, and $\W \sim Tt_{p}(\bmu,\bSigma,\nu;(\ap,\bp))$ stands for a $p$-variate doubly truncated Student's-$t$ distribution on $(\ap,\bp)\in\RR^p$.}

\begin{proposition}\label{prop2}
	If $\Y\sim Tt_p(\bmu,\bSigma,\nu;(\ap,\bp))$ then it follows that
	$$
	\EE\left[\left(\displaystyle\frac{\nu+p}{\nu+\delta(\Y)}\right)
	^r\Y^{(k)}\right] =c_p(\nu,r)\displaystyle
	\frac{T_p(\mathbf{a},\bp;\bmu,\bSigma^*,\nu+2r)}
	{T_{p}(\mathbf{a},\bp;\bmu,\bSigma,\nu)}\EE[\mathbf{W}^{(k)}],$$
	where 
	$$
	c_p(\nu,r)=\left(\displaystyle\frac{\nu+p}{\nu}\right)^r  \frac{\Gamma{\left(\frac{p+\nu}{2}\right)}\Gamma{\left(\frac{\nu+2r}{2}\right)}}
	{\Gamma\left(\frac{\nu}{2}\right)
		\Gamma{\left(\frac{p+\nu+2r}{2}\right)}} ,
	$$
	$\bSigma^*= {\nu}{}\bSigma/(\nu+2r)$ and $\nu+2r>0$, with $\mathbf{W}\sim Tt_p(\bmu,\bSigma^*,\nu+2r;(\ap,\bp))$.
\end{proposition}
\medskip

Notice that Proposition \ref{prop2} depends on  formulas for
$\EE[\mathbf{W}]$ and
$\EE[\mathbf{W}\mathbf{W}^{\top}]$, where $\mathbf{W} \sim
Tt_p(\bmu,\bSigma, \nu;(\ap,\bp))$. \CB{Having established the formula on the $k$-order moment of $\Y$, we provide an explicit formula for the
	conditional moments with respect to a two-component partition of $\Y$.}

\begin{proposition}\label{prop3} Let $\Y\sim
	{Tt}_p(\bmu,\bSigma,\nu;(\ap,\bp))$. Consider the partition $\Y^{\top} = (\Y^{\top}_1 ,
	\Y^{\top}_2 )$ with $\mathrm{dim}(\Y_1) = p_1$, $\mathrm{dim}(\Y_2) = p_2$, $p_1 + p_2
	= p$, and the corresponding partitions of  $\ap$, $\bp$, $\bmu$,
	and $\bSigma$. Then,
	$$
	\EE\left[ \left(\displaystyle\frac{\nu+p}{\nu+\delta(\Y)}\right)^r\Y_2^{(k)}\mid\Y_1\right] =\frac{d_p(p_1,\nu,r)}
	{(\nu+\delta(\Y_1))^r}
	\displaystyle
	\frac{T_{p_2}(\mathbf{a}_2,\mathbf{b}_2; \bmu_{2.1},\widetilde{\bSigma}^*_{22.1},\nu+p_1+2r)}
	{T_{p_2}(\ap_2,\bp_2;\bmu_{2.1},\widetilde{\bSigma}_{22.1},\nu+p_1)}
	\EE[\mathbf{W}_2^{(k)}],
	$$
	for $\nu+p_1+2r>0$, with $\delta(\Y_1) = \delta(\Y_1;\bmu_1,\bSigma_{11})$,
	$$
	\widetilde{\bSigma}^*_{22.1}=\left(\displaystyle\frac{\nu+\delta_1}
	{\nu+2r+p_1}\right)\bSigma_{22.1}, \quad
	\text{and}
	\quad
	d_p(p_1,\nu,r)=\left({\nu+p}\right)^r \frac{\Gamma{\left(\frac{p+\nu}{2}\right)}
		\Gamma{\left(\frac{p_1+\nu+2r}{2}\right)}}
	{\Gamma\left(\frac{p_1+\nu}{2}\right)
		\Gamma{\left(\frac{p+\nu+2r}{2}\right)}},
	$$
	where $\bSigma_{22.1}$ is defined as in proposition \ref{prop1}. Moreover, $\W_2 \sim \T t_{p_2}(\bmu_{2.1},\widetilde{\bSigma}^*_{22.1},\nu+p_1+2r;[\mathbf{a}_2,\mathbf{b}_2])$.
\end{proposition}

Observe that propositions \ref{prop2} and \ref{prop3} depend on  formulas for $\mathrm{E}[\mathbf{W}]$ and $\mathrm{E}[\mathbf{W}\mathbf{W}^{\top}]$, where $\mathbf{W} \sim  Tt_p(\bmu,\bSigma, \nu;\mathbb{A})$. The general formulas for these moments were derived efficiently by \cite{GalarzaCran}, for which we use the \verb"MomTrunc"  package in \verb"R".

\section{Review of the  Heckman selection normal model}\label{SLnModel}
Sample selection or missing data is common in applied research.  The SL model consists of a linear
equation for the outcome, and a Probit equation for the sample selection mechanism. The outcome equation is
\begin{equation}\label{1HS}
Y_{1i}=\x^{\top}_i\bbeta+\epsilon_{1i},
\end{equation}
and the sample selection mechanism is characterized by the following latent linear equation:
\begin{equation}\label{2HS}
Y_{2i}=\w^{\top}_i\bgamma+\epsilon_{2i},
\end{equation}
for $i = 1, \ldots, n$. The vectors $\bbeta\in {\mathbb R}^p$ and 
$\bgamma \in {\mathbb R}^q$ are unknown regression parameters. $\x^{\top}_i=(x_{i1},\ldots,x_{ip})$ and  $\w^{\top}_i=(w_{i1},\ldots,w_{iq})$ are known characteristics. The covariates in $\x_i$ and $\w_i$ may overlap with each other, and the exclusion restriction holds when at least
one of the elements of $\w_i$ are not in $\x_i$.  The indicator for sample selection is $C_i = I(Y_{2i} > 0)$. Let $V_{1i}$ be the observed outcome, we observe the
outcome $V_{1i}$ if and only if $C_i > 0$, i.e., $Y_{1i} = {V_{1i}}$ if $C_i = 1$, and $Y_{1i} = NA$ if $C_i = 0$, where $NA$ indicates missing data.

\cite{heckman1979} assumes independent $(\ind)$ bivariate normal distribution for the error terms (SLn), as follows:
\begin{center}
	\begin{eqnarray}\label{nerror}
		\begin{bmatrix}
			\epsilon_{1i}  \\
			\epsilon_{2i} \\
		\end{bmatrix}\ind N_{2}\left(  \begin{bmatrix}
			0 \\
			0 \\
		\end{bmatrix},
		\bSigma=\begin{bmatrix}
			\sigma^2 & \rho\sigma \\
			\rho\sigma & 1 \\
		\end{bmatrix}
		\right),
	\end{eqnarray}
\end{center}
where the second diagonal element of $\bSigma$ is fixed at 1 in order to achieve full identifiability.  The SLn model (\ref{1HS})-(\ref{nerror}) is known as "Type 2 tobit model" in the econometrics literature  and is sometimes also referred to as the “Heckman model.” Absence of selection effect ($\rho=0$) implies that the outcomes are missing at random, and the observed outcomes are representative for inference of the population given the observed covariates.

Under the bivariate normal assumption, the mean equation for the outcomes if the selected samples is
\begin{equation}\label{correH}
\EE\left[Y_{1i}|C_i=1,\x_i,\w_i\right]=\x^{\top}_i\bbeta+\rho\sigma \lambda(\w^{\top}_i{\bgamma}),
\end{equation}
where $\lambda(a)=\displaystyle\frac{\phi(a)}{\Phi(a)}$ is the inverse Mills ratio, $\phi(.)$ and $\Phi(.)$ denote the pdf and cdf of the standard normal distribution, respectively. Therefore, the SLn problem can be treated as a model misspecification problem, because the mean equation for the outcomes of the selected samples is a linear function $\x^{\top}_i\bbeta$
with a nonlinear correction term $\rho\sigma \lambda(\w^{\top}_i{\bgamma})$. Based on (\ref{correH}), \cite{heckman1979} proposed a two-step procedure by first fitting a Probit model of $C_i$ on $\w_i$ to obtain $\hat \bgamma$. At the second stage, $\bbeta$ and $\rho^*=\sigma\rho$ are estimated by least squares regression of $Y_{1i}$ (the observed counterpart) on $\x_i$ and $\hat \lambda=\displaystyle\frac{\phi(\w^{\top}_i{\hat \bgamma})}{\Phi(\w^{\top}_i{\hat \bgamma})}$. The consistent  estimators of $\rho$ and $\sigma$ can then be obtained from $\hat \rho^*$, least square residual variance, and average-predicted probabilities from the probit model. The two-step
procedure is less efficient than the ML estimation, but it is robust to the deviation of the joint normality of the error
terms. The ML estimates of the SLn model can be calculated by Newton-Raphson iteration or the EM algorithm as discussed by \cite{zhao2020new}. In the next subsection, we propose 
a slight modification to the EM-type algorithms proposed by \cite{zhao2020new}, wherein all the parameters are updated
(M-step) by considering the outcome $(Y_{1i})$ and sample selection ($Y_{2i}$) as missing data \citep{vaida2009fast,Matos11}. 

Ignoring censoring for the moment, suppose that we have observations on $n$ independent individuals
\begin{equation}
\Y_1, \ldots , \Y_n \ind N_2(\bmu_i,\bSigma), \label{modeleq}
\end{equation}
where {for each $i \in \{ 1, \ldots , n\}$}, $\Y_i=(Y_{1i},Y_{2i})^{\top}$ is the vector
of independent responses  for sample unit $i$, $$\bmu_i=\X_{ic}\bbeta_c,\,\,\,\,{\it with}\,\,\,\,\X_{ic}=\begin{bmatrix}
\x^{\top}_i & 0 \\
0 & \w^{\top}_i \\
\end{bmatrix},\,\,\,\bbeta_c=\begin{bmatrix}
\bbeta \\
\bgamma\\
\end{bmatrix}$$
and the dispersion matrix $\bSigma$ depends on an 
unknown parameter vector $(\sigma,\rho)$. 
We consider the
approach proposed by \cite{vaida2009fast} and \cite{Matos11} to represent the model belong to the structure of a 
censored linear model. Thus, the observed data for the
$i$th subject is given by $(\bV_i, C_i),$ where $\bV_i$ represents
the vector of censored readings  and $C_i=I_{\{Y_{2i}>0\}}$ is the censoring indicators. In other words,
\begin{equation}
Y_{1i}= V_{1i}, \mbox{ if } C_{i}=1\quad
\textrm{and}\quad Y_{1i}=V_{2i}=NA, \mbox{ if }
C_{i}=0,\label{CensL1}
\end{equation}
for all $i \in \{ 1, \ldots, n\}$.  Notice that $V_{2i}=NA$ is equivalent to write $-\infty<V_{2i}< \infty$.

\subsection{The likelihood function}\label{Likelihood_tMLC}
To obtain the likelihood function of the SLn model, first note that if $C_i=1$, then $Y_{1i}\sim N(\x^{\top}_i\bbeta,\sigma^2)$ and $Y_{2i}|Y_{1i}=V_{1i}\sim N(\mu_{c},\sigma^2_c)$, where

$$\mu_c=\w^{\top}_i\bgamma+\frac{\rho}{\sigma}(V_{1i}-\x^{\top}_i\bbeta),\,\,\sigma^2_c=(1-\rho^2).$$
Thus, the contribution in the likelihood is
$$f(Y_{1i}|\btheta)P(Y_{2i}>0|Y_{1i}=V_{1i})=\phi(V_{1i}|\x^{\top}_i\bbeta,\sigma^2)\Phi(\displaystyle\frac{\mu_{c}}{\sigma_c}).$$
If $C_i=0$, then the contribution in the likelihood is
$$P(Y_{2i}\leq 0)=\Phi(-\w^{\top}_i\bgamma).$$

Therefore, the likelihood function of $\btheta=(\bbeta^{\top},\bgamma^{\top},\sigma^2,\rho)^{\top}$ is
\begin{align}
L(\btheta\mid\bV,\bC)&=\prod_{i=1}^{n}\left[\phi(V_{1i}|\x^{\top}_i\bbeta,\sigma^2)\Phi(\displaystyle\frac{\mu_{c}}{\sigma_c})\right]^{C_i}\left[\Phi(-\w^{\top}_i\bgamma)\right]^{1-C_i},\label{equ8.1}
\end{align}
where   $\bV=(\bV_1,\ldots,\bV_n)$
and $\bC=(C_1,\ldots,C_n)$.  The log-likelihood function for the observed data is given by $\ell(\theta)=\ell(\btheta\mid\bV,\bC)=\ln L(\btheta\mid\bV,\bC)$,

\subsection{Parameter estimation via the EM algorithm}\label{EM-tlcm}

We describe in detail how to carry out ML estimation for the
SLn model. Let $\yp=(\yp^{\top}_1,\ldots,\yp^{\top}_n)^{\top}$,
$\bV=(\bV_1,\ldots,\bV_n)$ and $\bC=(C_1,\ldots,C_n)$, and
that we observe $(\textbf{V}_i,C_i)$ for the $i$th subject, $\ii$.  In
the estimation procedure, $\mathbf{y}$ is
treated as hypothetical missing data, and augmented with the
observed data set $(\bC,\bV)$, we have $\yp_c=(\bC,\bV,\yp)^{\top}$.
Hence, the EM-type algorithm is applied to the complete-data
log-likelihood function given by
$$\ell_c(\btheta|\mathbf{y}_c)=\sumas \ell_{ic}(\btheta),$$
where
\begin{equation*} 
 \ell_{ic}(\btheta)=-\frac{1}{2}\bigl\{ \ln|\bSigma|+(\yp_i-\bmu_i)^{\top}\bSigma^{-1}(\yp_i-\bmu_i)\bigr\} +c,
\end{equation*}
where  $c$ is  a constant that does not depend on $\btheta$, $\bmu_i=\X_{ic}\bbeta_c$ and $\bSigma$ as defined in (\ref{nerror}). Finally, the EM
algorithm for the SLn model can be summarized through the
following two steps.

\bigskip
\noindent \textbf{E-step:}
Given the current estimate  $\btheta=\widehat{\btheta}^{(k)}$ at the $k$th step of the algorithm, the
E-step provides the conditional expectation of the complete data
log-likelihood function
\begin{equation*}
Q(\btheta\mid\widehat{\btheta}^{(k)})=\mathrm{E} \Bigl[ \ell_c(\btheta|\yp_c )\mid\bV,\bC,\widehat{\btheta}^{(k)} \Bigr]
=\sumas{Q_i(\btheta\mid\widehat{\btheta}^{(k)})},\label{eq:Em:Q}
\end{equation*}
where
\begin{equation*}
Q_i(\btheta\mid\widehat{\btheta}^{(k)})=
Q_{i}(\bbeta,\bgamma,\sigma^2,\rho\mid\widehat{\btheta}^{(k)}) =
-\frac{1}{2}\ln{|\bSigma|}-\frac{1}{2}\tr\left[\left\{\widehat{\yp_i^{2}}^{(k)}-\widehat{\yp}^{(k)}_i{\bmu_i}^{\top} -{\bmu_i} \widehat{\yp}^{\top(k)}_i+{\bmu_i}{\bmu_i}^{\top}\right\}\bSigma^{-1}\right], \nonumber
\end{equation*}
with $\widehat{\yp}^{(k)}_i=\mathrm{E}[\displaystyle
\Y_i\mid\bV_i,C_i,\widehat{\btheta}^{(k)}]$ and
$\widehat{\yp_i^{2}}^{(k)}=\mathrm{E}[\displaystyle
\Y_i\Y_i^{\top}\mid\bV_i,C_i,\widehat{\btheta}^{(k)}]$ and   $\tr{(\A)}$ indicates the trace of the matrix $\A$.
Following \cite{Matos11}, for $C_i=1$ we have 
$$\widehat{\yp}^{(k)}_i=(V_{i}, \hat{w^c}_i)^{\top},\,\,\widehat{\yp_i^{2}}^{(k)}=\begin{bmatrix}
V^2_{1i} & V_{1i} \hat{w^c}_i\\
V^2_{1i}\hat{w^c}_i & \hat{w^{2c}}_i, \\
\end{bmatrix},$$
where $w^c_i=E[W_i|\btheta^{(k)}]$, $w^{2c}_i=E[W^2_i|\btheta^{(k)}]$, with $W_i\sim TN_1(\mu_c,\sigma^2_c;(0,\infty))$. For $C_i=0$
$$\widehat{\yp}^{(k)}_i=E[\bW_i|\btheta^{(k)}],\,\,\,\widehat{\yp^2}^{(k)}_i=E[\bW_i\bW^{\top}_i|\btheta^{(k)}],$$
where $\bW_i\sim TN_2(\bmu_i,\bSigma;{\mathbb A})$, with
\begin{equation} \label{hyper2}
\mathbb{A} = \{(x_1, x_2)\in \mathbb{R}^2: -\infty \leq x_1 \leq \infty, -\infty \leq x_2 \leq 0\}.
\end{equation}
 Here $TN_p(\bmu,\bSigma, \mathbb{ A})$ denotes the p-variate truncated normal distribution with location $\bmu$, scale matrix $\bSigma$ over the truncation region $\mathbb{A}$.\\
\bigskip
\noindent \textbf{M-step:} By the invariance
property of ML estimators, we use the parameter transformations $\psi = \sigma^2(1 -\rho^2)$  and $\rho^* = \rho\sigma$ in order to get closed form expression. Thus, 
in this step, $Q(\btheta\mid\widehat{\btheta}^{(k)})$ is conditionally
maximized with respect to $\btheta$ and a new estimate
$\widehat{\btheta}^{(k+1)}$ is obtained. Specifically, we have that
\begin{eqnarray}
\widehat{\bbeta_c}^{(k+1)}&=&\left(\sumas\X^{\top}_{ic}\widehat{\bSigma}^{(k)}\X_i \right)^{-1}\sumas
\X^{\top}_{ic}\widehat{\bSigma}^{(k)}\widehat{\yp}^{(k)}_i,\label{eq:beta_nMLCn}\\
\widehat{\psi}^{(k+1)}&=&\frac{1}{n}\sumas\left(\widehat\Gamma^{(k)}_{11i}-\widehat \rho^{*(k)}(\widehat \Gamma^{(k)}_{12i}+\widehat\Gamma^{(k)}_{21i})+\widehat \rho^{*2(k)} \widehat\Gamma^{(k)}_{22i}\right),\\
\widehat{\rho^*}^{(k+1)}&=&\frac{\sumas \left( \widehat \Gamma^{(k)}_{12i}+\widehat\Gamma^{(k)}_{21i}\right)}{2\sumas \widehat\Gamma^{(k)}_{22i}}, \label{eq:beta_nMLCn2} 
\end{eqnarray}
where $\widehat\Gamma^{(k)}_{kli}$ is the $kl$th element of the matrix  $\widehat\bGamma^{(k)}_{i}=\left\{\widehat{\yp_i^{2}}^{(k)}-\widehat{\yp}^{(k)}_i\widehat{\bmu_i}^{(k)\top}- \widehat{\bmu_i}^{(k)}(\widehat{\yp}^{(k)}_i)^\top+\widehat{\bmu_i}^{(k)}\widehat{\bmu_i}^{(k)\top}\right\}$. The algorithm is terminated when the relative distance between two successive evaluations of the log-likelihood defined in (\ref{equ8.1}) is less than a tolerance, i.e., {\small $|\ell(\widehat{\btheta}^{(k+1)}\mid\bV,\bC)/\ell(\widehat{\btheta}^{(k)}\mid\bV,\bC)-1|<\epsilon$}, for example, {\small $\epsilon=10^ {-6}$}.  Once converged, we can recover {\small $\widehat\sigma^2$} and {\small $\widehat\rho$} using the expressions
{\small\begin{equation*}\label{recover}
	\widehat\sigma^2=\widehat{\psi}+\widehat{\rho^*}^2
	\qquad\text{and}\qquad
	\widehat\rho = \displaystyle\frac{\widehat{\rho^*}}{\widehat{\sigma}}.
	\end{equation*}}
The initial values of the parameters for the EM algorithm are obtained from the 2-step procedure, which are obtained from the \verb"R" package \verb"sampleSelection" \citep{HenningsenCRAN}.

\subsection{Provision of standard errors}  \label{sec SE}

In this section, we describe how to obtain the standard errors of the ML estimates for the SLn model. We follow the information-based method exploited by \cite{basford1997standard} to compute the asymptotic covariance of the ML estimates. The empirical information matrix, according to \cite{meilijson1989fast}'s formula, is defined as
{\small\begin{eqnarray}\label{eq:IM}
	\bI_e(\btheta|\yp) = \sum_{i = 1}^{n} s(\yp_i|\btheta)s^\top (\yp_i|\btheta) - \frac{1}{n} S(\yp_i|\btheta)S^\top (\yp_i|\btheta),
	\end{eqnarray}}
where {\small$S(\yp_i|\btheta) = \sum_{i = 1}^{N} s(\yp_i|\btheta)$} and {\small$s(\yp_i|\btheta)$} is the empirical score function for the {\small$i$}th subject. It is noted from the result of \cite{louis1982finding} that the individual score can be determined as
{\small\begin{eqnarray}
	s(\yp_i|\btheta) = \E\left[ \left. \frac{\partial\ell_{ic}(\btheta)}{\partial \btheta}\right| \bV_i, C_i,\btheta\right].
	\end{eqnarray}}
Using the ML estimates {\small$\widehat{\btheta}$} in {\small$s(\yp_i|\btheta)$}, leads to {\small$S(\yp_i|\widehat{\btheta}) = 0$}, so from (\ref{eq:IM}) we have that
{\small \begin{eqnarray}\label{eq:oim}
	\bI_e(\widehat{\btheta}|\yp) = \sum_{i = 1}^{n} \widehat{\se}_i \widehat{\se}^\top_i,
	\end{eqnarray}}
where {\small$\widehat{\se}_i$} is an individual score vector given by {\small $\widehat{\se}_i = (\widehat{s}_{i,\bbeta_{c}}, \widehat{s}_{i,{\sigma}},\widehat{s}_{i,{\rho}})$. So, the expressions for the elements of {\small$\widehat{\se}_i$} are given by:

\begin{eqnarray*}
	\widehat{s}_{i,\bbeta_c} &=& \frac{1}{2}\X^{\top}_{ic}\bSigma^{-1}{\widehat \y_i}+\frac{1}{2}{\widehat \y}^{\top}_i\bSigma^{-1}\X_{ic}-\X^{\top}_{ic}\bSigma^{-1}\X_{ic}\bbeta_c,\\
	\widehat{s}_{i,\sigma} &=&  -\frac{1}{2}\tr(\bSigma^{-1}\B)+\frac{1}{2}\tr(\bGamma_i\bSigma^{-1}\B \Sigma^{-1}),
	\\
	\widehat{s}_{i,\rho} &=&  -\frac{1}{2}\tr(\bSigma^{-1}\D)+\frac{1}{2}\tr(\bGamma_i\bSigma^{-1}\D \bSigma^{-1}),  
		\end{eqnarray*}
where $\bGamma_i$ is as defined in (\ref{eq:beta_nMLCn})-(\ref{eq:beta_nMLCn2}), $\bbeta_c$ and $\X_{ic}$ as defined in (\ref{modeleq}), $\B=\begin{bmatrix}
2\sigma & \rho \\
\rho & 0 \\
\end{bmatrix}$, and $\D=\begin{bmatrix}
0 & \sigma \\
\sigma & 0 \\
\end{bmatrix}$.

The SLn model discussed in this section was criticized in the literature because of its sensitivity to the normality assumption. In the next section, we establish a new link between the SL model and the Student's-t distribution, called the Heckman selection-t model (SLt), as introduced by \cite{marchenko2012heckman}.

\section{The Heckman selection-t model} \label{SLtModel}

In order to accommodate for heavy-tailedness, \cite{marchenko2012heckman} proposed the SLt model, replacing the normal assumption of error terms in (\ref{nerror}) by a bivariate Student'-t distribution with an unknown number of degrees of freedom $\nu$:
\begin{center}
	\begin{eqnarray}\label{modeleqt}
		\begin{bmatrix}
			\epsilon_{1i}  \\
			\epsilon_{2i} \\
		\end{bmatrix}\sim t_{2}\left(  \begin{bmatrix}
			0 \\
			0 \\
		\end{bmatrix},
		\bSigma=\begin{bmatrix}
			\sigma^2 & \rho\sigma \\
			\rho\sigma & 1 \\
		\end{bmatrix},\nu
		\right).
	\end{eqnarray}
\end{center}
As in the SLn model, and ignoring censoring for the moment, we can represent the robust SLt model belong to a censored data framework, as follows:
\begin{equation*}
\Y_1, \ldots , \Y_n \ind t_{2}(\bmu_i,\bSigma,\nu), 
\end{equation*}
where {for each $i \in \{ 1, \ldots , n\}$}, $\Y_i=(Y_{i1},Y_{i2})^{\top}$ is the vector
of independent responses for sample unit $i$, $\bmu_i=\X_{ic}\bbeta_{c}$ with $\X_{ic}$ and $\bbeta_c$ as defined in (\ref{modeleq}). We consider the
approach proposed by \cite{vaida2009fast}, \cite{Matos11} and \cite{Matos.SINICA} to represent the model belong to the structure of a  censored linear model, where the observed data for the $i$th subject is given by $(\bV_i, C_i),$ where $\bV_i$ represents the vector of censored readings  and $C_i=I_{\{Y_{2i}>0\}}$ is the censoring indicators. The model defined in (\ref{1HS}), (\ref{2HS}), (\ref{CensL1}) and (\ref{modeleqt}) is henceforth called the SLt model.
\subsection{The likelihood function}\label{Likelihood_tMLC2}
To obtain the likelihood function of the SLt model, first  from Proposition \ref{prop1}  note that if $C_i=1$, then $Y_{1i}\sim t(\x^{\top}_i\bbeta,\sigma^2,\nu)$ and $Y_{2i}|Y_{1i}=V_{1i}\sim t(\mu_{ti},\sigma^2_{ti},\nu+1)$, where

$$\mu_{ti}=\w^{\top}_i\bgamma+\frac{\rho}{\sigma}(V_{1i}-\x^{\top}_i\bbeta),\,\,\sigma^2_{ti}=\frac{\nu+\delta(V_{1i})}{\nu+1}(1-\rho^2),$$
where $\delta(V_i)=\displaystyle\frac{(V_{1i}-\x^{\top}_i\bbeta)^2}{\sigma^2}$. Thus, the contribution in the likelihood function of $\btheta=(\bbeta^{\top},\bgamma^{\top},\sigma^2,\rho,\nu)^{\top}$, given the observed sample $(\bV, \C)$, is
$$f(Y_{1i}|\btheta)P(Y_{2i}>0|Y_{1i}=V_i)=t(V_{1i}|\x^{\top}_i\bbeta,\sigma^2,\nu)T(-\infty,0;-\mu_{ti},\sigma^2_{ti},\nu+1).$$

If $C_i=0$, then the contribution in the likelihood function is
$$P(Y_{2i}\leq 0)=T(-\infty,0;\w^{\top}_i\bgamma,1,\nu).$$

Therefore, the likelihood function of $\btheta=(\bbeta^{\top},\bgamma^{\top},\sigma^2,\rho,\nu)^{\top}$,  given the observed sample $(\bV, \C)$, is
\begin{align}
L(\btheta\mid\bV,\bC)&=\prod_{i=1}^{n}\left[t(V_{1i}|\x^{\top}_i\bbeta,\sigma^2,\nu)T(-\infty,0;-\mu_{ti},\sigma^2_{ti},\nu+1)\right]^{C_i}\left[T(-\infty,0;\w^{\top}_i\bgamma,1,\nu)\right]^{1-C_i},\label{equ8.2}
\end{align}
where   $\bV=(\bV_1,\ldots,\bV_n)$
and $\bC=(C_1,\ldots,C_n)$.  The log-likelihood function for the observed data is given by $\ell(\theta)=\ell(\btheta\mid\bV,\bC)=\ln L(\btheta\mid\bV,\bC)$, that is,

\begin{align}
\ell(\btheta)&=\sum_{i=1}^{n}\left[C_i\ln t(V_{1i}|\x^{\top}_i\bbeta,\sigma^2,\nu)+C_i\ln T(-\infty,0;-\mu_{ti},\sigma^2_{ti},\nu+1)\right]+\sum_{i=1}^{n}(1-C_i)\ln\left[ T(-\infty,0;\w^{\top}_i\bgamma,1,\nu)\right]. \label{equ8.3}
\end{align}

The ML estimate $\widehat \btheta$ of the vector of unknown parameters can be calculated by maximizing the log-likelihood given in (\ref{equ8.3}). There are many optimization procedures available in standard programs, such as the \verb"optim"  routine in \verb"R", which need only the original estimator function. A disadvantage of direct maximization of the log-likelihood function is that it may not converge unless good starting values are used. Thus, we also propose the EM algorithm for parameter estimation, which is stable and straightforward to implement since the iterations
converge monotonically and no second derivatives are required. Moreover, the EM estimates are quite insensitive to the starting values, as discussed by \cite{zhao2020new} regarding the SLn model.

\subsection{Parameter estimation via the EM algorithm}\label{EM-tlcm2}

Note first that by using the representation (\ref{stoNI1}) and ignoring censoring for the moment, we have that the distribution of $\Y_i$ can be hierarchically written  as 
\begin{align}
\Y_i \mid U_i=u_i \ind {\cal N}_{2}(\bmu_i,u_i^{-1}\bSigma), \quad
U_i \ind \mathcal{G}(\nu/2,\nu/2). \label{eqn tCMcomplete}
\end{align} 
Let $\yp=(\yp^{\top}_1,\ldots,\yp^{\top}_n)^{\top}$,
$\bV=(\bV_1,\ldots,\bV_n)$, $\bC=(C_1,\ldots,C_n)$, $\bu=(U_1,\ldots, U_n)$, and
that we observe $(\textbf{V}_i,C_i)$ for the $i$th subject.  In
the estimation procedure, $\mathbf{y}$ and $\bu$ are
treated as hypothetical missing data, and augmented with the
observed data set we have
$\yp_c=(\bC,\bV,\yp,\bu)^{\top}$ .
Hence, the EM-type algorithm is applied to the complete-data
log-likelihood function given by
$$\ell_c(\btheta|\yp_c)=\sumas \ell_{ic}(\btheta),$$
where
\begin{equation*} 
\ell_{ic}(\btheta)=-\frac{1}{2}\bigl\{ \ln|\bSigma|+u_i(\yp_i-\bmu_i)^{\top}\bSigma^{-1}(\yp_i-\bmu_i)\bigr\} +
\ln h(u_i\mid\nu)+c,
\end{equation*}
where  $c$ is  a constant that does not depend on $\btheta$ and $h(u_i\mid\nu)$
is the $\mbox{Gamma}(\nu/2,\nu/2)$ pdf. The EM
algorithm for the SLt model can be summarized through the
following two steps.

\bigskip
\noindent \textbf{E-step:}
Given the current estimate  $\btheta=\widehat{\btheta}^{(k)}$ at the $k$th step of the algorithm, the
E-step provides the conditional expectation of the complete data
log-likelihood function
\begin{equation*}
Q(\btheta\mid\widehat{\btheta}^{(k)})=\mathrm{E} \Bigl[ \ell_c(\btheta|\yp_c)\mid\bV,\bC,\widehat{\btheta}^{(k)} \Bigr]
=\sumas{Q_i(\btheta\mid\widehat{\btheta}^{(k)})},
\end{equation*}
where
\begin{equation*}
Q_i(\btheta\mid\widehat{\btheta}^{(k)})=
Q_{i}(\bbeta^{\top},\bgamma^{\top},\sigma^2,\rho,\nu\mid\widehat{\btheta}^{(k)}) =
-\frac{1}{2}\ln{|\bSigma|}-\frac{1}{2}\tr\left[\left\{\widehat{u\yp_i^{2}}^{(k)}-\widehat{u\yp}^{(k)}_i{\bmu_i}^{\top} -{\bmu_i} (\widehat{u\yp}^{(k)}_i)^{\top}+\widehat{u}^{(k)}_i{\bmu_i}{\bmu_i}^{\top}\right\}\bSigma^{-1}\right], \nonumber
\end{equation*}
with $\widehat{u\yp}^{(k)}_i=\mathrm{E}[\displaystyle
U_i\Y_i\mid\bV_i,\bC_i,\widehat{\btheta}^{(k)}]$,
$\widehat{u\yp_i^{2}}^{(k)}=\mathrm{E}[\displaystyle
U_i\Y_i\Y_i^{\top}\mid\bV_i,\bC_i,\widehat{\btheta}^{(k)}]$ and
$\widehat{u}^{(k)}_i=\mathrm{E}[\displaystyle
U_i\mid\bV_i,\bC_i,\widehat{\btheta}^{(k)}]$. Note that
$\widehat\kappa^{(k)}_i =\mathrm{E}\left[\ln h(U_i\mid\nu)\mid\bV,\bC,\widehat{\btheta}^{(k)}\right]$ is analytically intractable. Instead, we avoid the calculation of $\widehat\kappa^{(k)}_i$
by performing the CML-step for updating $\nu$.  As in the normal case, we use the parameter transformations $\psi = \sigma^2(1 -\rho^2)$  and $\rho^* = \rho\sigma$ in order to get closed form expression in the M-Step.

\bigskip
\noindent \textbf{M-step:}
In this step, $Q(\btheta\mid\widehat{\btheta}^{(k)})$ is conditionally
maximized with respect to $\bbeta_c,\sigma^2,\rho$ and a new estimate
$\widehat{\bbeta}^{(k+1)}_c,\widehat{\bsigma}^{2(k+1)},\widehat{\rho}^{(k+1)}$ is obtained. Specifically, we have that
\begin{eqnarray}
\widehat{\bbeta}^{(k+1)}_c&=&\left(\sumas\widehat{u}^{(k)}_i\X^{\top}_{ic}\bSigma^{(k)}\X_{ic}\right)^{-1}\sumas
\X^{\top}_{ic}\bSigma^{(k)}\widehat{u\yp}^{(k)}_i,\,\,\,\,\widehat{\bmu}^{(k+1)}_i=\X_{ic}\widehat{\bbeta}^{(k+1)}_c\label{eq:beta_tMLC0}\\
\widehat{\psi}^{(k+1)}&=&\frac{1}{n}\sumas\left(\widehat\Gamma^{(k)}_{11i}-\widehat \rho^{*(k)}(\widehat \Gamma^{(k)}_{12i}+\widehat\Gamma^{(k)}_{21i})+\widehat \rho^{*2(k)} \widehat\Gamma^{(k)}_{22i}\right),\\
\widehat{\rho^*}^{(k+1)}&=&\frac{\sumas \left( \widehat \Gamma^{(k)}_{12i}+\widehat\Gamma^{(k)}_{21i}\right)}{2\sumas \widehat\Gamma^{(k)}_{22i}}, \label{eq:beta_tMLCn2}\\
\widehat{\sigma}^{2(k+1)}&=&\widehat{\psi}^{(k+1)}+\widehat{\rho^*}^{2(k+1)}
	\qquad\text{and}\qquad\qquad
	\widehat\rho^{(k+1)} = \displaystyle\frac{\widehat{\rho}^{*(k+1)}}{\widehat{\sigma}^{(k+1)}}\label{eq:rho_tMLC}.
\end{eqnarray}
where $\widehat\Gamma^{(k)}_{kli}$ is the $kl$th element of the matrix  $\widehat\bGamma^{(k)}_{i}=\left\{\widehat{u\yp_i^{2}}^{(k)}-\widehat{u\yp}^{(k)}_i\widehat{\bmu_i}^{(k)\top}- \widehat{\bmu_i}^{(k)}(\widehat{\yp}^{(k)}_i)^\top+\widehat{u}^{(k)}_i\widehat{\bmu_i}^{(k)}\widehat{\bmu_i}^{(k)\top}\right\}.$\\
\noindent \textbf{CLM-step:} Update $\widehat \nu^{(k+1)}$ by maximizing the actual marginal
log-likelihood function, obtaining
\begin{eqnarray*}
\widehat{\nu}^{(k+1)}&=& argmax_\nu \left\{\sum_{i=1}^n C_i\ln \left[t(V_{1i}|\x^{\top}_i\widehat\bbeta^{(k+1)},\widehat\sigma^{2(k+1)},\nu)T(-\infty,0;-\widehat\mu^{(k+1)}_{ti},\widehat\sigma^{2(k+1)}_{ti},\nu+1)\right]
\right.\\ && \left.+ \sum_{i=1}^n(1-C_i)
\ln T(-\infty,0;\w^{\top}_i\widehat\bgamma^{(k+1)},1,\nu)\right\} .
\end{eqnarray*}

The algorithm is iterated until a suitable convergence rule is
satisfied. It is important to stress that,  from
{Eqs.~}~(\ref{eq:beta_tMLC0}) - (\ref{eq:rho_tMLC}), the E-step
reduces to the computation of $\widehat{u\yp_i^{2}}$,
$\widehat{u\yp}_i$, and $\widehat{u}_i$. To compute these expected values, first observe that they can be written in terms of $\mathrm{E}(U_i \mid \Y_i)$, where $\Y_i\sim t_{p}(\bmu,\bSigma,\nu)$ -- see the definition of $U_i$ in (\ref{eqn tCMcomplete}).  

For example, we have that $\widehat{u}_i= \mathrm{E}\left[ \mathrm{E}(U_i\mid \Y_i)\mid \bV_i,\bC_i,\widehat{\btheta}^{(k)}\right] $. It is straightforward to prove that $\mathrm{E}[U_i \mid \Y_i]=(\nu+1)/(\nu+\delta)$, where $\delta=(\Y_i-\bmu_i)^{\top}\bSigma^{-1}(\Y_i-\bmu_i)$. Then, we can use  Propositions \ref{prop2} and  \ref{prop3} to obtain closed form expressions as follows:
\begin{itemize}

	\item[1.] If $C_i=0$, from Proposition 3, we have
	\begin{eqnarray*}
		\widehat{u\yp_i^{2}}^{(k)}&=&\mathrm{E}[\displaystyle U_i\Y_i\Y_i^{\top}\mid\bV_i,C_i,\widehat{\btheta}^{(k)}]=\widehat{\varphi}^{(k)}(\bV_i)\widehat{\bw}_i^{2 ^{c(k)}},\,\\
		\widehat{u\yp}^{(k)}_i&=&\mathrm{E}[\displaystyle U_i\Y_i\mid\bV_i,C_i,\widehat{\btheta}^{(k)}]=\widehat{\varphi}^{(k)}(\bV_i)\widehat{\bw}_i^{c(k)},\,\, \\
		\widehat{u}^{(k)}_i&=&\mathrm{E}[\displaystyle
		U_i\mid\bV_i,C_i,\widehat{\btheta}^{(k)}]=\widehat{\varphi}^{(k)}(\bV_i),
	\end{eqnarray*}
	where 
	$$
	\widehat{\varphi}^{(k)}(\bV_i)=\frac{T_{2}((-\infty,-\infty),(\infty,0);\widehat{\bmu_i}^{(k)}, \quad \widehat{\bSigma}^{*(k)},\nu+2)}{T_{2}((-\infty,-\infty),(\infty,0); \widehat{\bmu_i}^{(k)}, \quad \widehat{\bSigma}^{(k)},\nu)}, 
	\quad \widehat{\bw}_i^{c(k)}=\mathrm{E}[\mathbf{W}_i\mid\widehat{\btheta}^{(k)}], \quad \widehat{\bw}_i^{2 ^{c(k)}}=\mathrm{E}[\bW_i \bW_i^\top\mid\widehat{\btheta}^{(k)}],$$
	$$
	\mathbf{W}_i\sim
	Tt_{2}(\widehat{\bmu_i}^{(k)},\widehat{\bSigma}^{*(k)},\nu+2;\mathbb{A}), \quad
	\widehat{\bSigma}^{*(k)}=\displaystyle\frac{\nu}{\nu+2}\widehat{\bSigma}^{(k)},
	$$
	with $\mathbb{A}_i$ as defined in (\ref{hyper2}).  To  compute  $\mathrm{E}[\mathbf{W}_i]$ and
	$\mathrm{E}[\mathbf{W}_i\mathbf{W}_i^{\top}]$ we use  the \verb"R" package \verb"MomTrunc"  \citep{GalarzaCran}. 
	
	\item[2.] If $C_i=1$,
	then from Proposition \ref{prop3}, we have that
	\begin{align*}
	\widehat{u\yp_i^{2}}^{(k)}&=\mathrm{E}[\displaystyle
	U_i\Y_i\Y_i^{\top}\mid Y_{1i},\bV_i,C_i,\widehat{\btheta}^{(k)}]=\left(\begin{array}{cc}
	V^2_{1i} \widehat{u}^{(k)}_i & \widehat{u}^{(k)}_i\widehat{{w}}^{c(k)\top}_iV_{1i} \\
	\widehat{u}^{(k)}_i\widehat{{w}}^{c(k)}_iV_{1i}  &\widehat{u}^{(k)}_i \widehat{w}_i^{2 ^{c(k)}}\\
	\end{array}\right ),\\
	\widehat{u\yp}^{(k)}_i&=\mathrm{E}[\displaystyle U_i\Y_i\mid Y_{1i},\bV_i,\bC_i,\widehat{\btheta}^{(k)}]=\mathrm{vec}(V_{1i}\widehat{u}^{(k)}_i,\widehat{u}^{(k)}_i\widehat{{w}}^{c(k)}_i),\,\, \\
	\widehat{u}^{(k)}_i&=\mathrm{E}[\displaystyle
	U_i\mid Y_{1i},\bV_i,\bC_i,\widehat{\btheta}^{(k)}]=\left\{ \frac{\nu^{(k)}+1}{\nu^{(k)}+\widehat{\delta}^{(k)}(V_{1i})}\right\}\displaystyle\frac{T(0,\infty;\widehat{\mu}^{(k)}_{ti},\widetilde{{S}}_i^{(k)},\nu^{(k)}+3)}{T(0,\infty;\widehat{\mu}^{(k)}_{ti},\hat \sigma^{2(k)}_{ti},\nu^{(k)}+1)},
	\end{align*}
	where 
	$${\hat \mu}^{(k)}_{ti}=\w^{\top}_i{{\hat \bgamma}^{(k)}}+\frac{{\hat \rho}^{(k)}}{{\hat \sigma}^{(k)}}(V_i-\x^{\top}_i{\hat\bbeta}^{(k)}),\,\,
	\widetilde{{S}}_i^{(k)}=\left\{ \displaystyle\frac{\nu^{(k)}+\widehat{\delta}^{(k)}(V_{1i})}{\nu^{(k)}+3}\right\}(1-\hat{\rho^2}^{(k)}), \quad \widehat{\delta}^{(k)}(V_{1i})=\frac{(V_i-\x^{\top}_i{\hat\bbeta}^{(k)})}{\hat \sigma^{(k)}}, 
	$$
	$$\hat \sigma^{2(k)}_{ti}=\frac{\nu^{(k)}+\hat \delta^{(k)}(V_{1i})}{\nu^{(k)}+1}(1-\hat\rho^{2(k)}),$$
	 with  
	${W}_i\sim
	Tt(\widehat{\mu}^{(k)}_{ti},\widetilde{{S}}_i^{(k)},\nu+3;[0,\infty))$.  Again, to  compute  $\widehat{{w}}^{c(k)}_i=\mathrm{E}[{W}_i]$ and
	$\widehat{w}_i^{2 ^{c(k)}}=\mathrm{E}[{W}^2_i]$ we use the \verb"R" package \verb"MomTrunc".   
\end{itemize}

\subsection{Provision of standard errors}  \label{sec SE1}
To obtain the standard errors of the ML estimates for the SLt model, we use the same strategy discussed in Subsection  \ref{sec SE} for the SLn model. It is important to stress that the standard error of $\nu$ depends heavily on the calculation of $\widehat\kappa_i =\mathrm{E}\left[ \ln h(U_i\mid\nu)\mid\bV,\bC,\widehat{\btheta}\right]$, which relies
on computationally intensive Monte Carlo integration. In
our analysis, we focus solely on comparing the standard errors
of $\bbeta_c=(\bbeta^{\top},\bgamma^{\top})^{\top},\,\, \sigma$ and $\rho$. So, the expressions for the elements of {\small$\widehat{\se}_i$} are given by:
	
	\begin{eqnarray*}
		\widehat{s}_{i,\bbeta_c} &=& \frac{1}{2}\X^{\top}_{ic}\bSigma^{-1}{\widehat{u\yp_i}}+\frac{1}{2}{\widehat{u\yp_i}}^{\top}\bSigma^{-1}\X_{ic}-\widehat{u}_i\X^{\top}_{ic}\bSigma^{-1}\X_{ic}\bbeta_c,\\
		\widehat{s}_{i,\sigma} &=&  -\frac{1}{2}\tr(\bSigma^{-1}\B)+\frac{1}{2}\tr(\bGamma_i\bSigma^{-1}\B \Sigma^{-1}),
		\\
		\widehat{s}_{i,\rho} &=&  -\frac{1}{2}\tr(\bSigma^{-1}\D)+\frac{1}{2}\tr(\bGamma_i\bSigma^{-1}\D \bSigma^{-1}),  
	\end{eqnarray*}
	
	where $\bGamma_i$ as in (\ref{eq:beta_tMLC0})-(\ref{eq:rho_tMLC}), $\B$ and $\D$ like defined in Subsection \ref{sec SE}.
	
\begin{table}[!htb]
\caption{{\bf Simulation study 1.} Mean estimates (EM), mean standard errors (SE) and Monte Carlo standard error (MC SE) of the $500$ Monte Carlo replicates for the data generated from the Normal distribution. The mean AIC and BIC are reported for the SLn and SLt models.}
	\label{tab:sim_normal}
\centering
\begin{tabular}{cccccccc}
  \toprule
Sample Size & & \multicolumn{3}{c}{SLn} & \multicolumn{3}{c}{SLt} \\
  \cmidrule{3-5} \cmidrule{6-8}    
 & TRUE & EM & SE & MC SE & EM & SE & MC SE \\ 
  \midrule
250 & $\beta_0=1.000$ & 1.017 & 0.147 & 0.130 & 1.020 & 0.145 & 0.128 \\ 
  & $\beta_1=0.500$ & 0.497 & 0.133 & 0.134 & 0.495 & 0.133 & 0.133 \\ 
  & $\gamma_0=0.674$ & 0.689 & 0.092 & 0.094 & 0.697 & 0.094 & 0.096 \\ 
  & $\gamma_1=0.300$ & 0.316 & 0.157 & 0.157 & 0.320 & 0.159 & 0.160 \\ 
  & $\gamma_2=-0.500$ & -0.512 & 0.096 & 0.095 & -0.521 & 0.099 & 0.097 \\ 
  & $\sigma^2=1.000$ & 1.004 & 0.085 & 0.081 & 0.983 & 0.089 & 0.081 \\ 
  & $\rho=0.600$ & 0.550 & 0.270 & 0.216 & 0.547 & 0.264 & 0.218 \\ 
  & $\nu=+\infty$ &  &  &  & 111.899 &  &  \\
  & AIC & \multicolumn{3}{c}{736.672} & \multicolumn{3}{c}{736.455}\\ 
  & BIC & \multicolumn{3}{c}{740.193} & \multicolumn{3}{c}{739.977}\\\midrule
500 & $\beta_0=1.000$ & 1.006 & 0.090 & 0.092 & 1.009 & 0.090 & 0.091 \\ 
  & $\beta_1=0.500$ & 0.505 & 0.088 & 0.092 & 0.503 & 0.088 & 0.092 \\ 
  & $\gamma_0=0.674$ & 0.681 & 0.066 & 0.066 & 0.687 & 0.067 & 0.067 \\ 
  & $\gamma_1=0.300$ & 0.304 & 0.108 & 0.111 & 0.308 & 0.109 & 0.112 \\ 
  & $\gamma_2=-0.500$ & -0.504 & 0.064 & 0.068 & -0.511 & 0.065 & 0.069 \\
  & $\sigma^2=1.000$ & 1.000 & 0.057 & 0.057 & 0.985 & 0.059 & 0.058 \\ 
  & $\rho=0.600$ & 0.584 & 0.153 & 0.152 & 0.580 & 0.153 & 0.153 \\ 
  & $\nu=+\infty$ &  &  &  & 118.592 &  &  \\ 
    & AIC & \multicolumn{3}{c}{1480.419} & \multicolumn{3}{c}{1480.244}\\ 
  & BIC & \multicolumn{3}{c}{1484.633} & \multicolumn{3}{c}{1484.459}\\\midrule
1000   & $\beta_0=1.000$ & 1.004 & 0.064 & 0.064 & 1.007 & 0.062 & 0.064 \\ 
  & $\beta_1=0.500$ & 0.500 & 0.065 & 0.065 & 0.499 & 0.065 & 0.065 \\ 
  & $\gamma_0=0.674$ & 0.680 & 0.045 & 0.046 & 0.685 & 0.046 & 0.047 \\ 
  & $\gamma_1=0.300$ & 0.304 & 0.078 & 0.078 & 0.307 & 0.079 & 0.078 \\ 
  & $\gamma_2=-0.500$ & -0.503 & 0.047 & 0.049 & -0.508 & 0.048 & 0.049 \\
  & $\sigma^2=1.000$ & 1.000 & 0.041 & 0.041 & 0.987 & 0.042 & 0.041 \\ 
  & $\rho=0.600$ & 0.593 & 0.107 & 0.106 & 0.589 & 0.105 & 0.107 \\ 
  & $\nu=+\infty$ &  &  &  & 117.861 &  &  \\ 
    & AIC & \multicolumn{3}{c}{2974.585} & \multicolumn{3}{c}{2974.495}\\ 
  & BIC & \multicolumn{3}{c}{2979.493} & \multicolumn{3}{c}{2979.403}\\
   \bottomrule
\end{tabular}
\end{table}

\section{Simulation study} \label{secSim}

In this section, we present three simulations studies. In the first one, we study the finite sample properties of the EM estimates as well as how the robustness of the parameter estimation for the SLt model performs in comparison to model the SLn in the presence of model misspecification. The other two studies focus on the vulnerability of the SLn model when data have heavy-tail and varying the correlation parameter $\rho$ and percentage of missing. For each scenario $500$ Monte Carlo samples were generated.

\subsection{Finite sample properties}
\label{sec:s1}
The initial goal of this simulation is to show the capacity of the proposed EM algorithm to correctly recover the parameters in SL models. Moreover, we test the robustness of the SL alternatives generating data from the Slash distribution.
\begin{table}[ht]
\caption{{\bf Simulation study 1.} Mean estimates (EM), mean standard errors (SE) and Monte Carlo standard error (MC SE) of the $500$ Monte Carlo replicates for the data generated from the Student-t with $4$ degrees of freedom. The mean AIC and BIC are reported for the SLn and SLt models.}
	\label{tab:sim_t}
\centering
\begin{tabular}{cccccccc}
  \toprule
Sample Size & & \multicolumn{3}{c}{SLn} & \multicolumn{3}{c}{SLt} \\
  \cmidrule{3-5} \cmidrule{6-8}    
 & TRUE & EM & SE & MC SE & EM & SE & MC SE \\ 
  \midrule
250 & $\beta_0=1.000$ &  0.964 & 0.260 & 0.176 & 1.011 & 0.146 & 0.131 \\ 
  & $\beta_1=0.500$ & 0.515 & 0.189 & 0.193 & 0.498 & 0.152 & 0.152 \\  
  & $\gamma_0=0.741$ & 0.689 & 0.092 & 0.094 & 0.697 & 0.094 & 0.096 \\ 
  & $\gamma_1=0.300$ & 0.316 & 0.157 & 0.157 & 0.320 & 0.159 & 0.160 \\ 
  & $\gamma_2=-0.500$ &  -0.410 & 0.104 & 0.087 & -0.511 & 0.115 & 0.109 \\ 
  & $\sigma^2=1.000$ & 1.406 & 0.219 & 0.079 & 1.024 & 0.113 & 0.089 \\ 
  & $\rho=0.600$ & 0.596 & 0.379 & 0.161 & 0.573 & 0.234 & 0.191 \\ 
  & $\nu=4$ &  &  &  & 6.904 &  &  \\    
  & AIC & \multicolumn{3}{c}{846.417} & \multicolumn{3}{c}{824.338}\\ 
  & BIC & \multicolumn{3}{c}{849.938} & \multicolumn{3}{c}{827.859}\\\midrule
500 & $\beta_0=1.000$ &  0.916 & 0.162 & 0.115 & 1.001 & 0.092 & 0.091 \\ 
  & $\beta_1=0.500$ & 0.536 & 0.126 & 0.130 & 0.506 & 0.101 & 0.104 \\ 
  & $\gamma_0=0.741$ & 0.643 & 0.066 & 0.065 & 0.747 & 0.081 & 0.078 \\ 
  & $\gamma_1=0.300$ & 0.255 & 0.105 & 0.108 & 0.307 & 0.127 & 0.128 \\ 
  & $\gamma_2=-0.500$ & -0.395 & 0.074 & 0.060 & -0.503 & 0.079 & 0.077 \\ 
  & $\sigma^2=1.000$ & 1.414 & 0.173 & 0.051 & 1.012 & 0.076 & 0.063 \\ 
  & $\rho=0.600$ & 0.678 & 0.240 & 0.089 & 0.596 & 0.147 & 0.133 \\ 
  & $\nu=4$ &  &  &  & 4.442 &  &  \\   
  & AIC & \multicolumn{3}{c}{1704.538} & \multicolumn{3}{c}{1658.626}\\ 
  & BIC & \multicolumn{3}{c}{849.938} & \multicolumn{3}{c}{827.859}\\\midrule
1000   & $\beta_0=1.000$ &  0.894 & 0.104 & 0.076 & 1.005 & 0.061 & 0.063 \\
  & $\beta_1=0.500$ & 0.533 & 0.089 & 0.091 & 0.500 & 0.076 & 0.074 \\
  & $\gamma_0=0.741$ & 0.642 & 0.045 & 0.045 & 0.746 & 0.057 & 0.055 \\
  & $\gamma_1=0.300$ & 0.248 & 0.075 & 0.075 & 0.303 & 0.090 & 0.089 \\
  & $\gamma_2=-0.500$ & -0.389 & 0.051 & 0.043 & -0.502 & 0.055 & 0.056 \\ 
  & $\sigma^2=1.000$ & 1.421 & 0.116 & 0.034 & 1.004 & 0.053 & 0.044 \\
  & $\rho=0.600$ & 0.723 & 0.142 & 0.047 & 0.596 & 0.098 & 0.095 \\ 
  & $\nu=4$ &  &  &  & 4.166 &  &  \\ 
  & AIC & \multicolumn{3}{c}{3427.196} & \multicolumn{3}{c}{3331.052}\\ 
  & BIC & \multicolumn{3}{c}{3432.103} & \multicolumn{3}{c}{3335.960}\\
   \bottomrule
\end{tabular}
\end{table}

Thus, for this study, data are generated from the normal, Student's-t, and slash distributions, respectively, with increasing sample sizes, say, $n=250$, $500$, and $1000$. The components of $\w_i^\top = (1, w_{i1}, w_{i2})$ are generated from a Uniform$(-1,1)$ and Normal$(0,1)$, respectively. To consider the exclusion restriction we let $\x_i^\top = (1, w_{i1})$, $\ii$. For all scenarios $\gamma = (\beta_{0d},0.3,-0.5)$ and $\beta = (1,0.5)$. The $\beta_{0d}$ is selected as the $-25\%$ quantile times $\sigma^2$ of the generating distribution in order to guarantee an average missing rate of $25\%$. Finally, the scale parameters are set at $\sigma^2 = 1$ and $\rho = 0.6$.

\begin{table}[ht]
\caption{{\bf Simulation study 1.} Mean estimates (EM), mean standard errors (SE) and Monte Carlo standard error (MC SE) of the $500$ Monte Carlo replicates for the data generated from the Slash with $1.43$ degrees of freedom. The mean AIC and BIC are reported for the SLn and SLt models.}
	\label{tab:sim_sl}
\centering
\begin{tabular}{cccccccc}
  \toprule
Sample Size & & \multicolumn{3}{c}{SLn} & \multicolumn{3}{c}{SLt} \\
  \cmidrule{3-5} \cmidrule{6-8}    
 & TRUE & EM & SE & MC SE & EM & SE & MC SE \\ 
  \midrule
250 & $\beta_0=1.000$ & 0.989 & 0.392 & 0.235 & 1.020 & 0.197 & 0.167 \\ 
  & $\beta_1=0.500$ & 0.510 & 0.238 & 0.248 & 0.492 & 0.181 & 0.184 \\ 
  & $\gamma_0=0.925$ & 0.651 & 0.100 & 0.092 & 0.751 & 0.116 & 0.108 \\ 
  & $\gamma_1=0.300$ & 0.214 & 0.152 & 0.153 & 0.255 & 0.175 & 0.176 \\ 
  & $\gamma_2=-0.500$ & -0.327 & 0.099 & 0.083 & -0.412 & 0.107 & 0.103 \\ 
  & $\sigma^2=1.000$ & 1.781 & 0.395 & 0.091 & 1.272 & 0.145 & 0.113 \\ 
  & $\rho=0.600$ & 0.543 & 0.466 & 0.181 & 0.561 & 0.271 & 0.204 \\ 
  & $\nu=1.43$ &  &  &  & 10.273 &  &  \\ 
  & AIC & \multicolumn{3}{c}{936.927} & \multicolumn{3}{c}{907.287}\\ 
  & BIC & \multicolumn{3}{c}{940.448} & \multicolumn{3}{c}{910.808}\\\midrule
500 & $\beta_0=1.000$ & 0.904 & 0.292 & 0.163 & 1.016 & 0.126 & 0.116 \\ 
  & $\beta_1=0.500$ & 0.534 & 0.162 & 0.178 & 0.501 & 0.124 & 0.126 \\ 
  & $\gamma_0=0.925$ & 0.645 & 0.073 & 0.064 & 0.747 & 0.081 & 0.076 \\ 
  & $\gamma_1=0.300$ & 0.208 & 0.103 & 0.108 & 0.250 & 0.122 & 0.126 \\ 
  & $\gamma_2=-0.500$ & -0.315 & 0.068 & 0.058 & -0.408 & 0.073 & 0.074 \\ 
  & $\sigma^2=1.000$ & 1.807 & 0.490 & 0.057 & 1.246 & 0.093 & 0.080 \\ 
  & $\rho=0.600$ & 0.648 & 0.336 & 0.097 & 0.574 & 0.169 & 0.149 \\ 
  & $\nu=1.43$ &  &  &  & 5.226 &  &  \\
  & AIC & \multicolumn{3}{c}{1888.094} & \multicolumn{3}{c}{1821.601}\\ 
  & BIC & \multicolumn{3}{c}{1892.309} & \multicolumn{3}{c}{1825.815}\\\midrule
1000   & $\beta_0=1.000$ & 0.866 & 0.225 & 0.095 & 1.014 & 0.083 & 0.080 \\ 
  & $\beta_1=0.500$ & 0.532 & 0.120 & 0.119 & 0.495 & 0.092 & 0.089 \\ 
  & $\gamma_0=0.925$ & 0.643 & 0.054 & 0.045 & 0.750 & 0.057 & 0.054 \\ 
  & $\gamma_1=0.300$ & 0.203 & 0.080 & 0.075 & 0.250 & 0.093 & 0.088 \\ 
  & $\gamma_2=-0.500$ & -0.309 & 0.057 & 0.041 & -0.409 & 0.055 & 0.053 \\ 
  & $\sigma^2=1.000$ & 1.816 & 0.263 & 0.035 & 1.236 & 0.067 & 0.056 \\ 
  & $\rho=0.600$ & 0.701 & 0.256 & 0.046 & 0.583 & 0.107 & 0.104 \\ 
  & $\nu=1.43$ &  &  &  & 4.415 &  &  \\ 
  & AIC & \multicolumn{3}{c}{3803.225} & \multicolumn{3}{c}{3661.727}\\ 
  & BIC & \multicolumn{3}{c}{3808.132} & \multicolumn{3}{c}{3666.635}\\
   \bottomrule
\end{tabular}
\end{table}

From Table~\ref{tab:sim_normal} we can see that both SLn e SLt models recover closely the original values of the parameters for all sample sizes, but being more precise as the sample size $n$ increases. Also, it is possible to see that the estimated standard errors for the parameters are very close to the Monte Carlo standard deviations. The model selection criterion does not differentiate the fit, which is expected since the SLn model is a especial case of the SLt model when $\nu = +\infty$ and $\nu$ is estimated in high values for the SLt model.

Now, under the same specification of the parameters values, we generate data from a Student's-t with $\nu=4$. The results are given in Table~\ref{tab:sim_t}. From this table we can see that the SLn model does not perform well when data is generated from a Student's-t distribution with $\nu=4$. While the SLt fit seems to be adequate and improves as the sample sizes increase, the SLn model present bias in the estimation of the parameters that increase as sample sizes increase. Although this may seem counter-intuitive it is not, with a bigger sample size more data is observed at the tail, due to the fat tail of the Studen-t distribution, and as expected affect the SLn model estimation. The only parameter that seems not to suffer under the SLn model is $\beta_1$. We can see that as $n$ increases and more observation in the tails are available, the estimation of $\nu$ improves drastically. Also we can see that the model selection criterion (AIC and BIC) are in favor of the SLt model, this is, also expected since the SLt is the true generating model and the SLn does not provide an adequate fit for the heavy-tailed data.

To study the robustness of the SLn and SLt model under model misspecification, data are now generated from the slash distribution with heavy tails (degrees of freedom $1.43$) keeping the same specification of the other parameters values. The results are presented in Table~\ref{tab:sim_sl}. The model selection criteria are in favor of the SLt model, as expected, indicating that under model misspecification it provides a more apropriate fit.

As can be seen, the estimates of $\bbeta$ is better than for $\bgamma$ for the SLn and SLt models. However, the bias for the SLn model is larger than the SLt model, especially, when the sample sizes increase. Moreover, the estimates for the SLt model are more stable for the different sample sizes. For example, it is clear that as sample sizes increase the bias estimation for $\rho$ under the SLn model increases drastically while the SLt model does not suffer significant variation. As sample size increase we can see that the estimate of $\nu$ reduces, indicating that is necessary a heavy-tailed model to provide an appropriate fit. Finally, the standard error estimates for the SLt model is stable under all scenarios. 

\subsection{Dependence variation}
\label{sec:s2}

In this section, we set the values of the parameters as in the previous section and generate data from the SLt model with $\nu=4$. To study the effect of the dependence parameter, we vary  $\rho$ from mild, $\rho=0.20$, to strong, $\rho=0.80$, dependence by a step of $0.20$. 
\begin{figure}
\centering
\begin{subfigure}
  \centering
  \includegraphics[width=.28\linewidth]{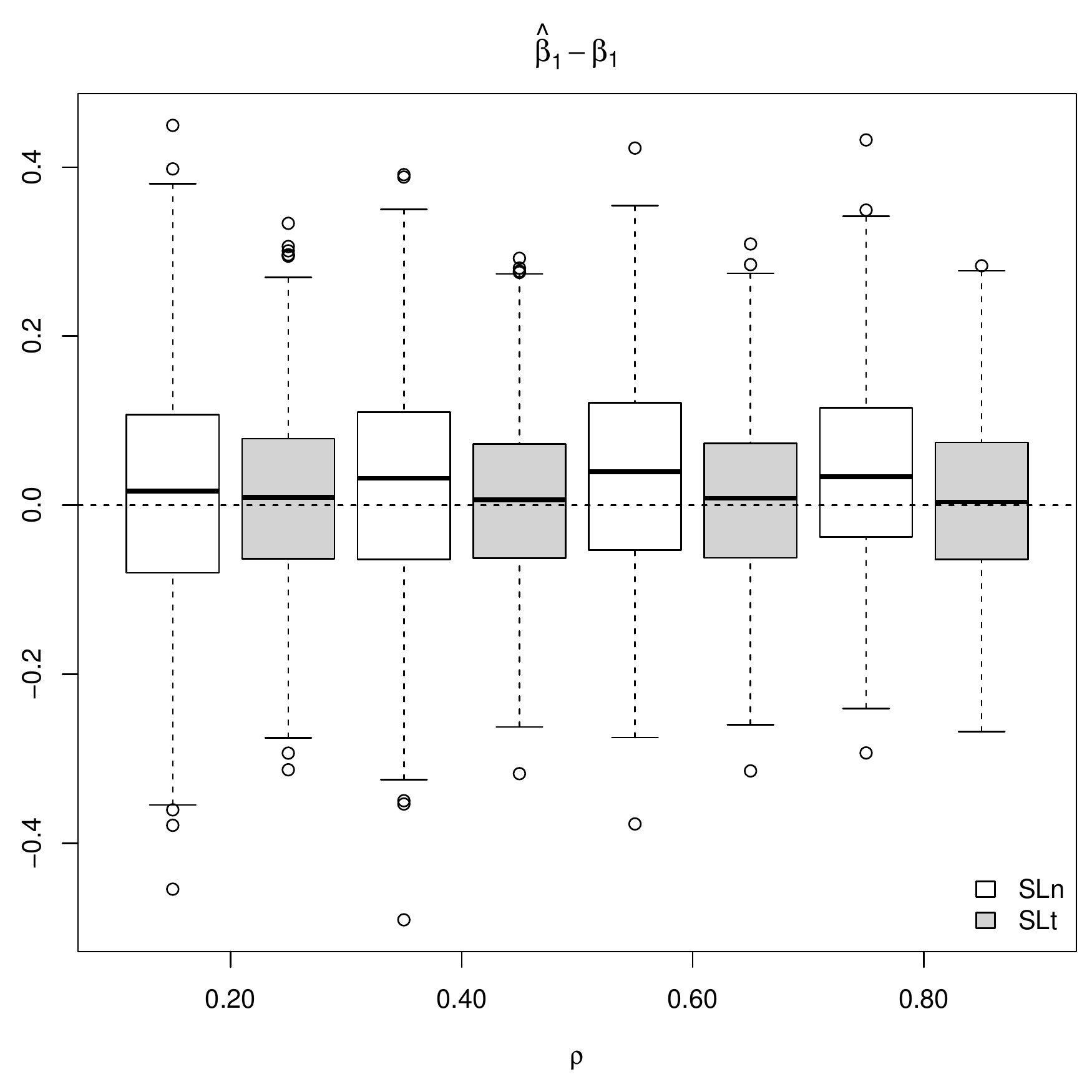}
\end{subfigure}%
\begin{subfigure}
  \centering
  \includegraphics[width=.28\linewidth]{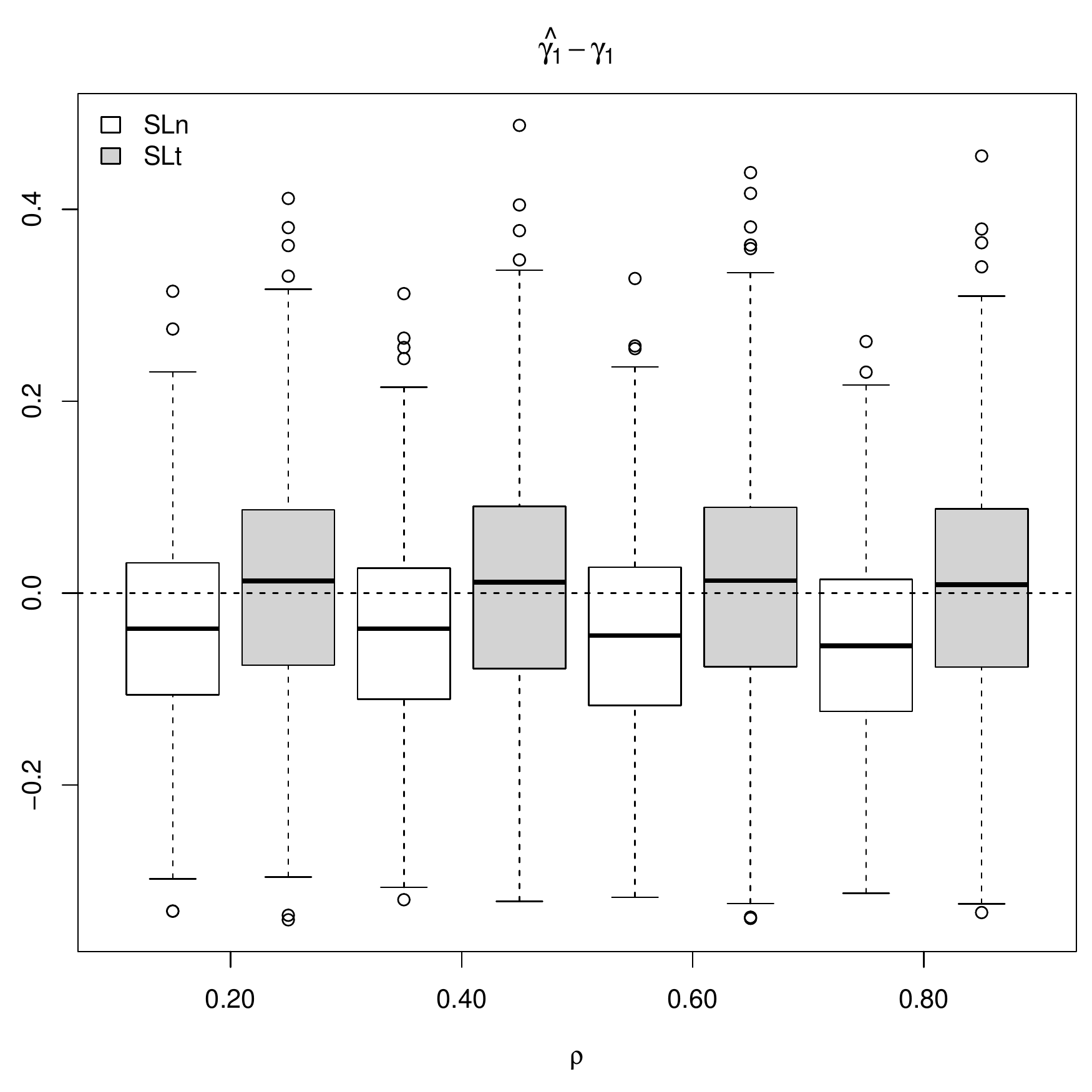}
\end{subfigure}
\begin{subfigure}
  \centering
  \includegraphics[width=.28\linewidth]{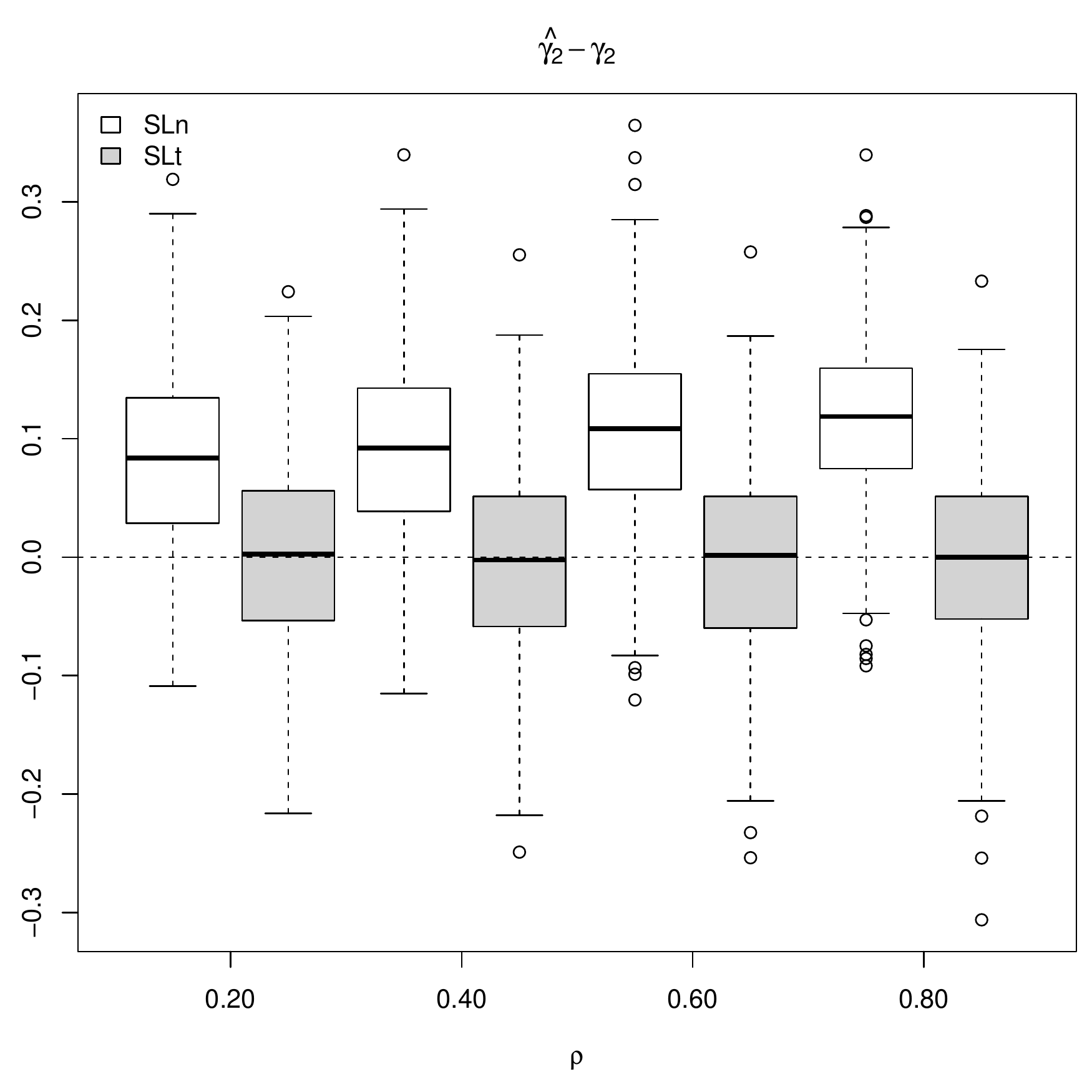}
\end{subfigure}\\
\begin{subfigure}
  \centering
  \includegraphics[width=.28\linewidth]{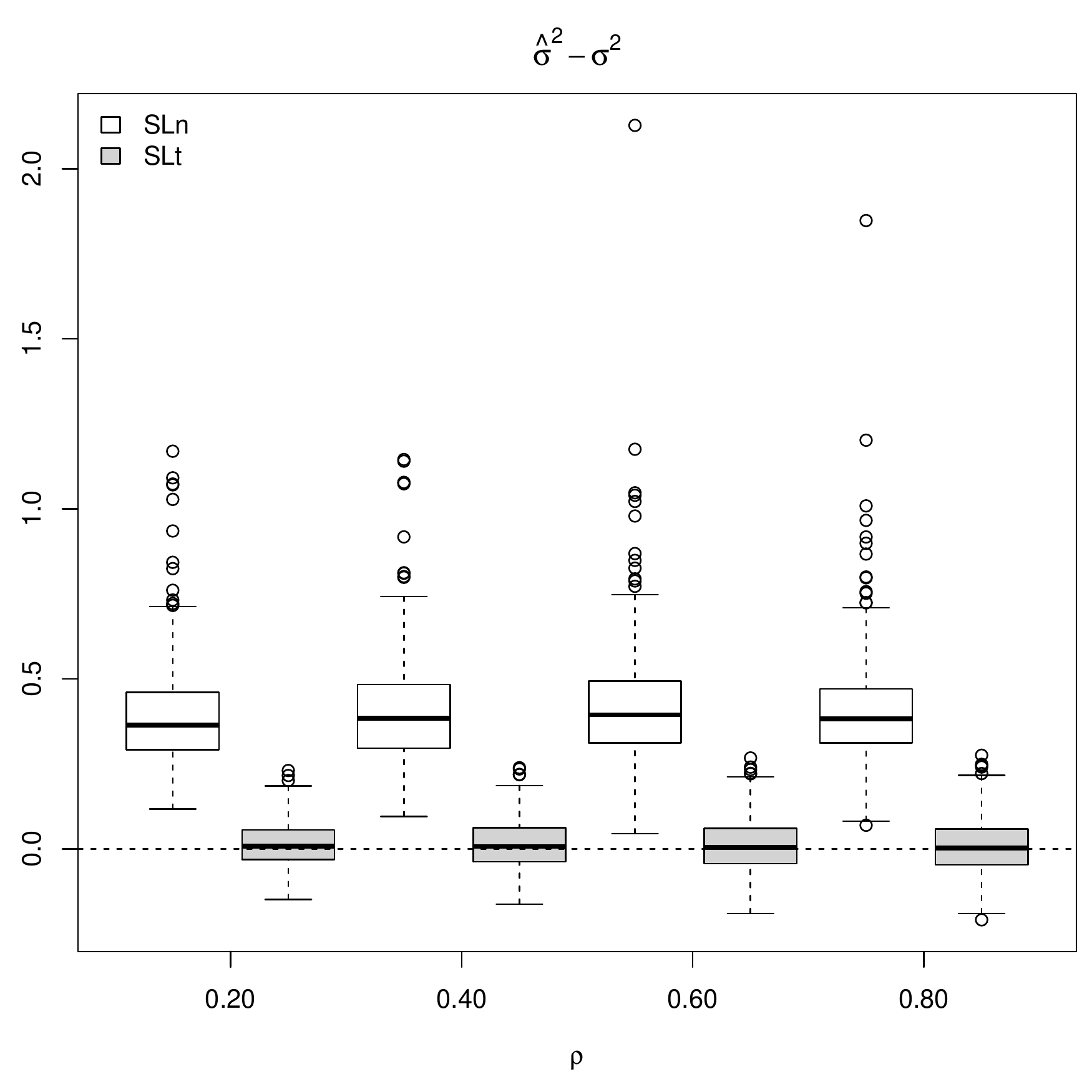}
\end{subfigure}
\begin{subfigure}
  \centering
  \includegraphics[width=.28\linewidth]{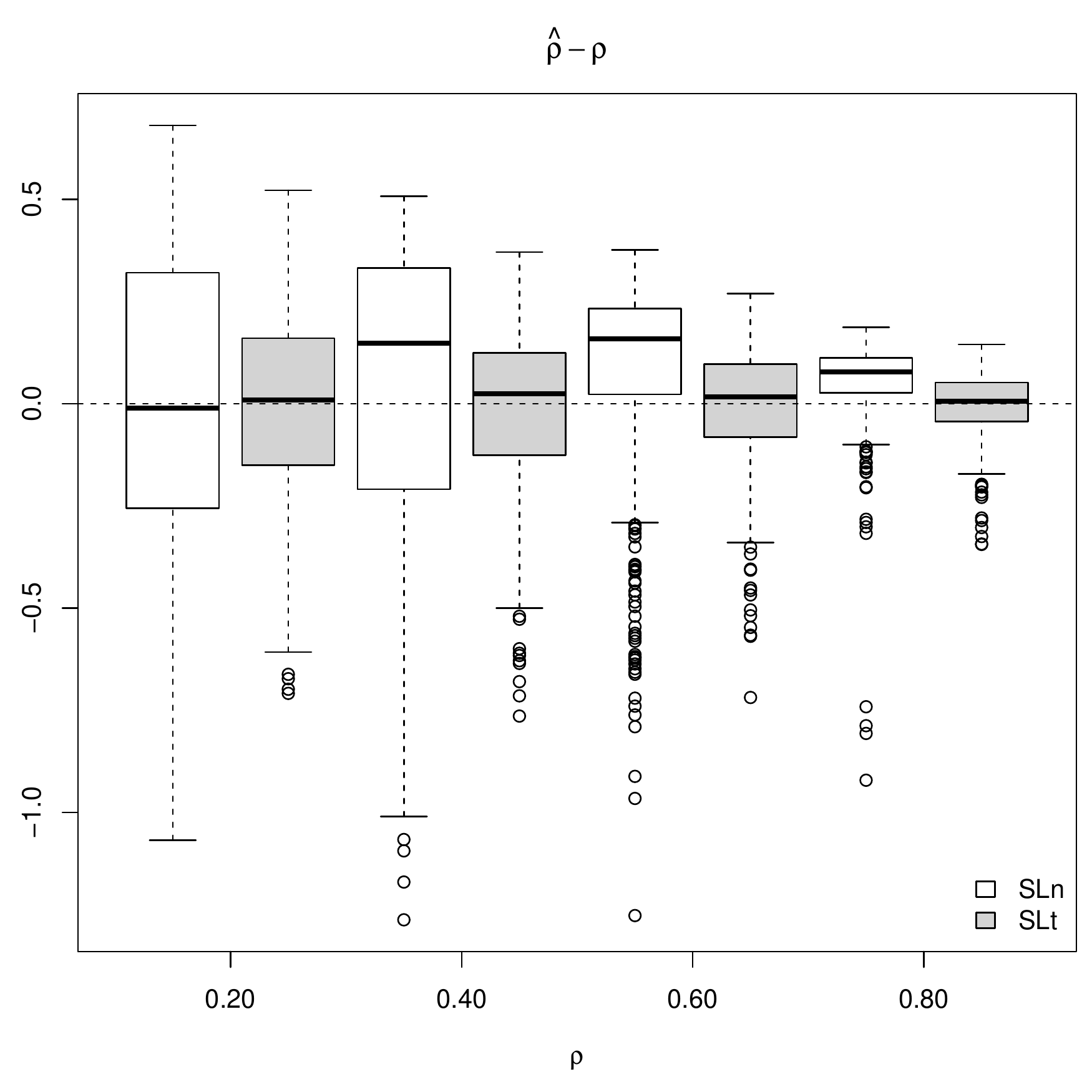}
\end{subfigure}
\caption{Show the boxplots of the estimated minus the true parameter for the SLn (white) and SLt (gray) for the different dependence values, $\rho=(0.20,0,40,0.60,0.80)$. }
\label{fig:rho}
\end{figure}

Figure~\ref{fig:rho} show the boxplots of the Monte Carlo fits of the SLn model (in white) and SLt model (in gray) for the $500$ Monte Carlo replicates. From this figure we can see that the SLn model provides large bias estimates with a larger variance. However, the SLt model provides accurate estimates for all parameters, except for the dependence parameter $\rho$ where the variability reduces as the dependence gets bigger. Although the SLn has a larger bias, the reduction in the estimates variability observed for the SLt model is also observed for this model.

\subsection{Missing variation}
\label{sec:s3}
To understand the effects of missing values in the model we vary the average percentage of missing from $10\%$, $25\%$, and $50\%$. We kept the other configuration of the SLt generating model from the previous section. Figure~\ref{fig:cens} shows the boxplots of the centered estimates of the parameters. As previously observed the SLt model have very stable results while the SLn have lager bias for all parameters. An interesting observation is that the variance for $\bgamma$ are larger for the SLt model with lower missing values than for largers ones.
\begin{figure}
\centering
\begin{subfigure}
  \centering
  \includegraphics[width=.28\linewidth]{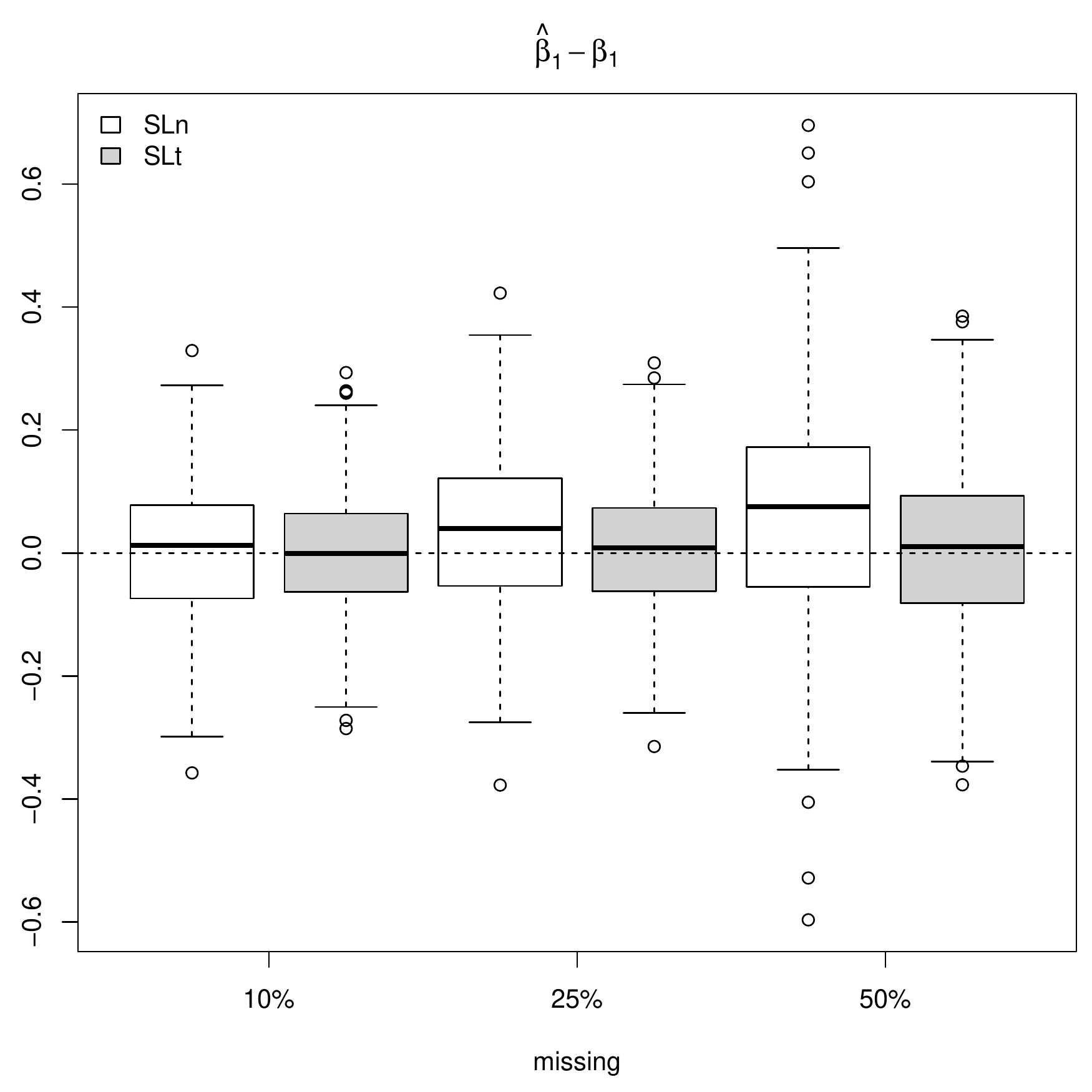}
\end{subfigure}%
\begin{subfigure}
  \centering
  \includegraphics[width=.28\linewidth]{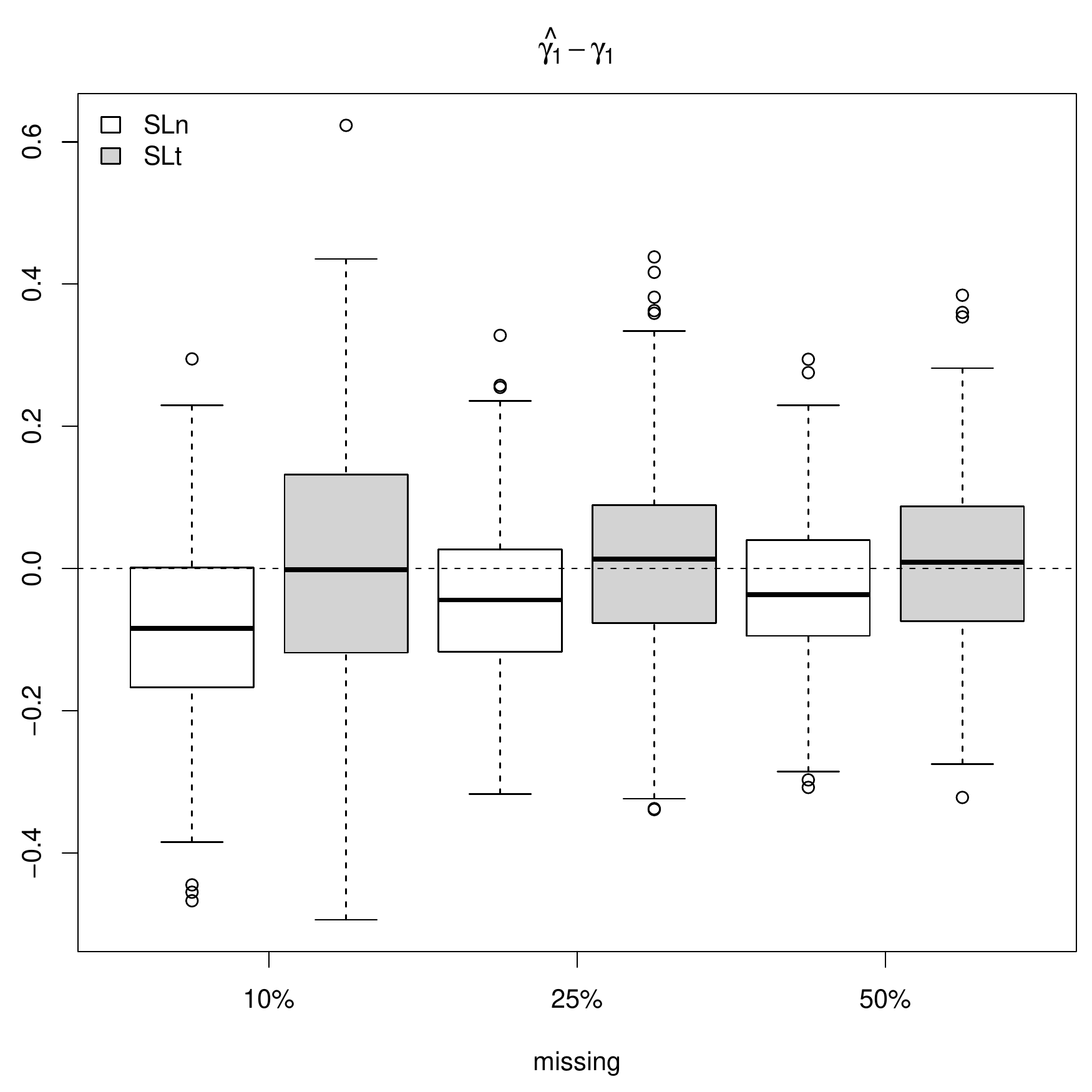}
\end{subfigure}
\begin{subfigure}
  \centering
  \includegraphics[width=.28\linewidth]{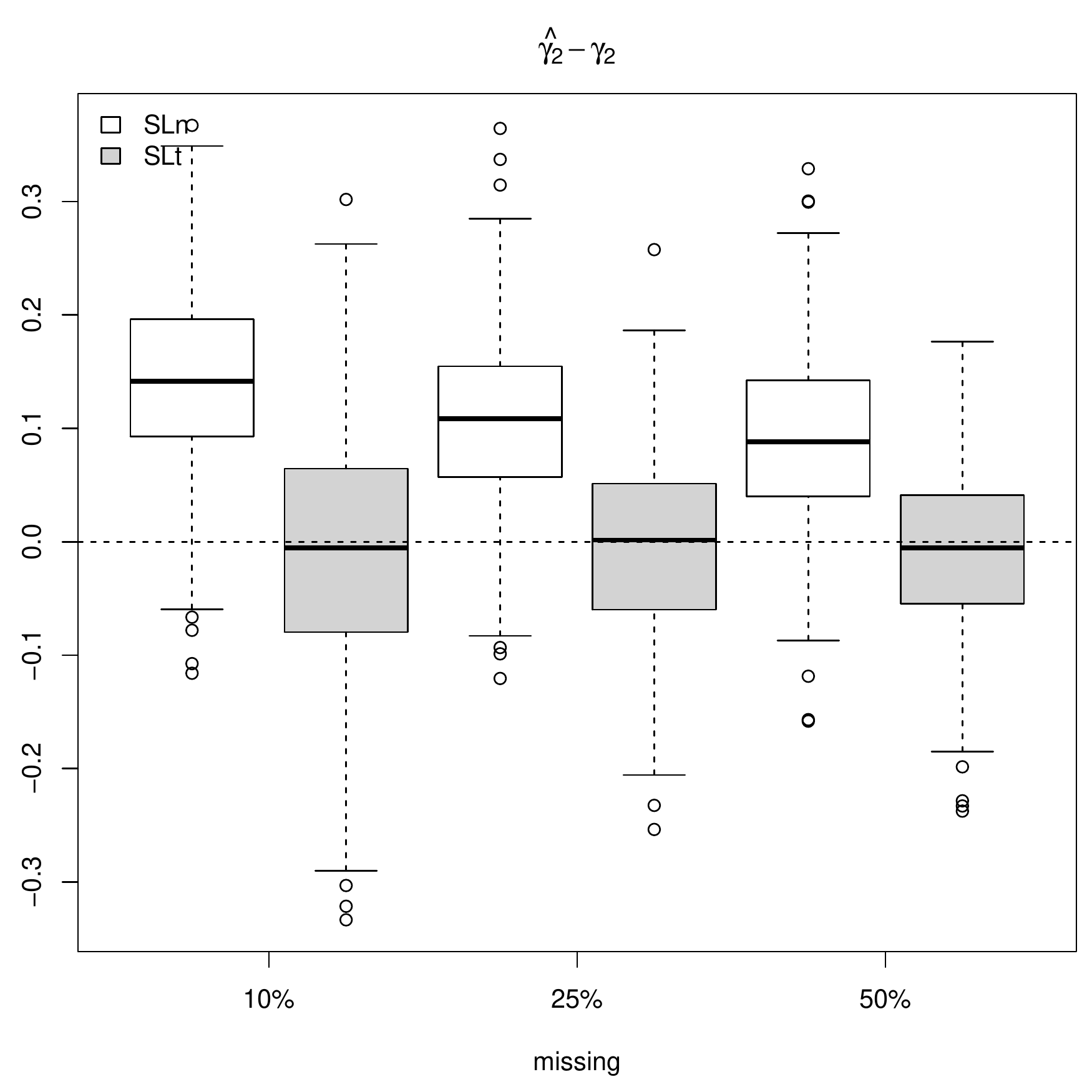}
\end{subfigure}\\
\begin{subfigure}
  \centering
  \includegraphics[width=.28\linewidth]{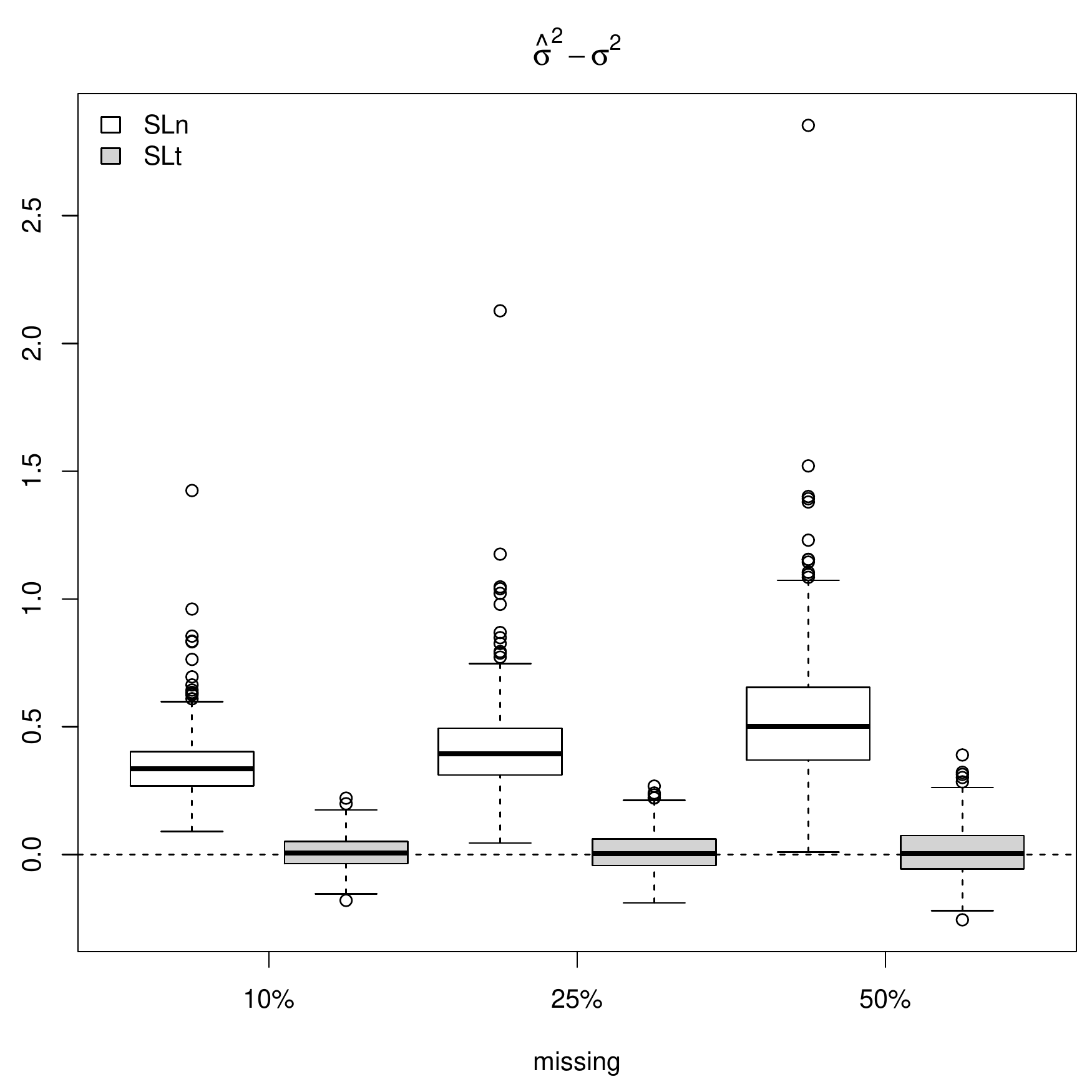}
\end{subfigure}
\begin{subfigure}
  \centering
  \includegraphics[width=.28\linewidth]{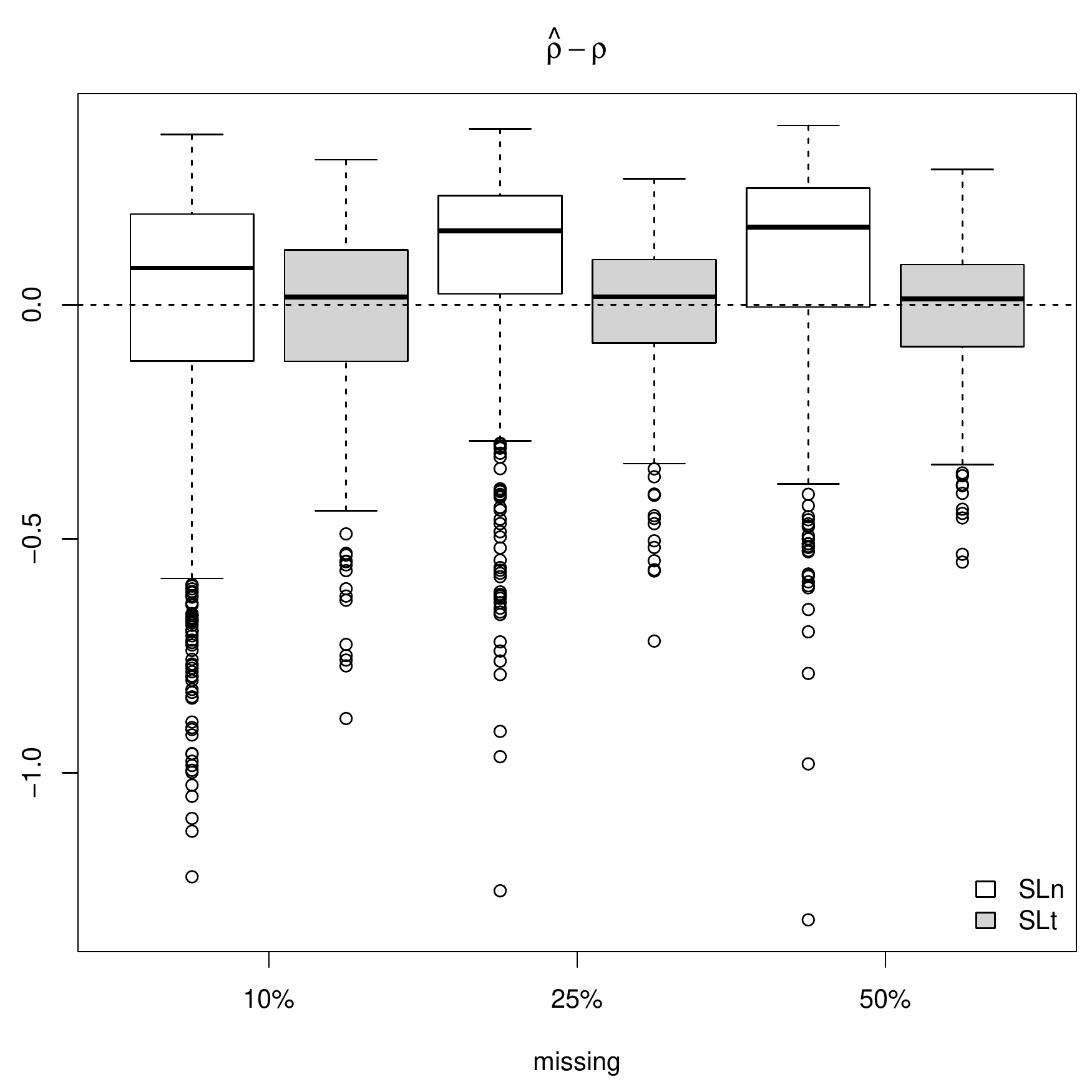}
\end{subfigure}
\caption{Show the boxplots of the estimated minus the true parameter for the SLn (white) and SLt (gray) for the different missing values, $10\%$, $25\%$, and $50\%$. }
\label{fig:cens}
\end{figure}

Overall, it is clear that the proposed algorithms provide good point estimates as well as adequate standard errors estimation (Section~\ref{sec:s1}) for both SLn and SLt models. Moreover, we verified that the SLt model is more robust than the SLn model for different variation on the characteristics of the generating model, as dependence (Section~\ref{sec:s2}), missing (Section~\ref{sec:s3}), and model misspecification (Section~\ref{sec:s1}).  

\section{Application}\label{secApp}
We illustrate the proposed algorithms with the analysis of two real data sets. The analysis was performed using the \verb"R" package \verb"HeckmanEM"  available on Github (\url{https://github.com/marcosop/HeckmanEM}). 

\subsection{Ambulatory expenditures}
The first application concerns a study of ambulatory expenditures. The data are taken from \cite{cameron2009microeconometrics}, which
were re-analyzed by \cite{marchenko2012heckman} using  ML estimation procedure in \verb"Stata". More recently, the data were also revisited by \cite{ding2014bayesian} using Bayesian procedures.  

In our analysis, we choose the same set of covariates as \cite{marchenko2012heckman},  we choose the  $\ln$ of ambulatory  expenditures
($ambexp$) as the outcome variable. The covariates in the outcome equation are $\xp = (1, age, female, educ, blhisp,
totchr, ins)$, including age, gender, education status, ethnicity, number of chronic diseases and insurance status, respectively. The exclusion restriction assumption holds by including the income variable into the selection equation, i.e., $\w = (\xp, income)$. The dataset contains 3328 observations and there are 526 missing
values of $ambexp$. More details about the data can be found
in Chap. 16 of \cite{cameron2009microeconometrics}. 

The estimation results for the SLn model and the SLt model are presented in Table \ref{tab:expediture}. As expected, we find similar results to those presented by \cite{marchenko2012heckman} and \cite{ding2014bayesian}. As noted  by these authors, under the SLn model, the 95\% confidence interval of $\rho$ contain zero $(-0.560;0.300)$, which indicates weak evidence of the SL bias. However, under the SLt model, the 95\% confidence interval of $\rho$  does not contain zero $(-0.369;-0.275)$, which suggests the existence of SL effect.

In order to identify atypical observations and/or model
misspecification, we use the martingale-type residuals,
as discussed by \cite{massuia2015influence} \citep[see also,][]{garay2016nonlinear} in the context of
for censored models. We propose to work with the $r_{MT_{i}}$ residual, which is given by: $$ r_{MT_{i}} =
\textrm{sign}(r_{M_{i}})\sqrt{-2[r_{M_{i}} + C_{i}\ln(C_{i} -
	r_{M_{i}})]}, \ \ i = 1, \ldots, n,$$ where $r_{M_{i}} = C_{i} +
\log(S(y_{i};\widehat{\btheta}))$ is the martingale residual, with
$C_{i} = 0,1$ as defined in (\ref{CensL1}), $sign(r_{M_{i}})$
denoting the sign of $r_{M_{i}}$ and $S(y_{i},\widehat{\btheta})=P_{\widehat{\btheta}}(Y_{2i}\leq0)$ evaluated in the ML estimates under the SLn model or the SLt model. A more detailed account of the martingale residual can be found in \cite{therneau1990martingale}.

The normal probability plot of the $r_{M_{i}}$ residuals with generated envelopes is presented in Figure \ref{fig:envelopes}. We observe that the SLt  model
fit the data better than the SLn model, since, in that case, there are fewer observations which lie outside the envelopes.  Moreover, Table \ref{tab:expediture} (bottom) presents some model selection criteria together with the values of
the log-likelihood.  The AIC and BIC values indicate that the SLt model with heavy tails presents a better fit than the SLn model, due to the departure of the data from normality.

\begin{figure}[!ht]
	\begin{center}
		\centering \hspace {1cm}\centering \\
		\includegraphics[scale=0.5]{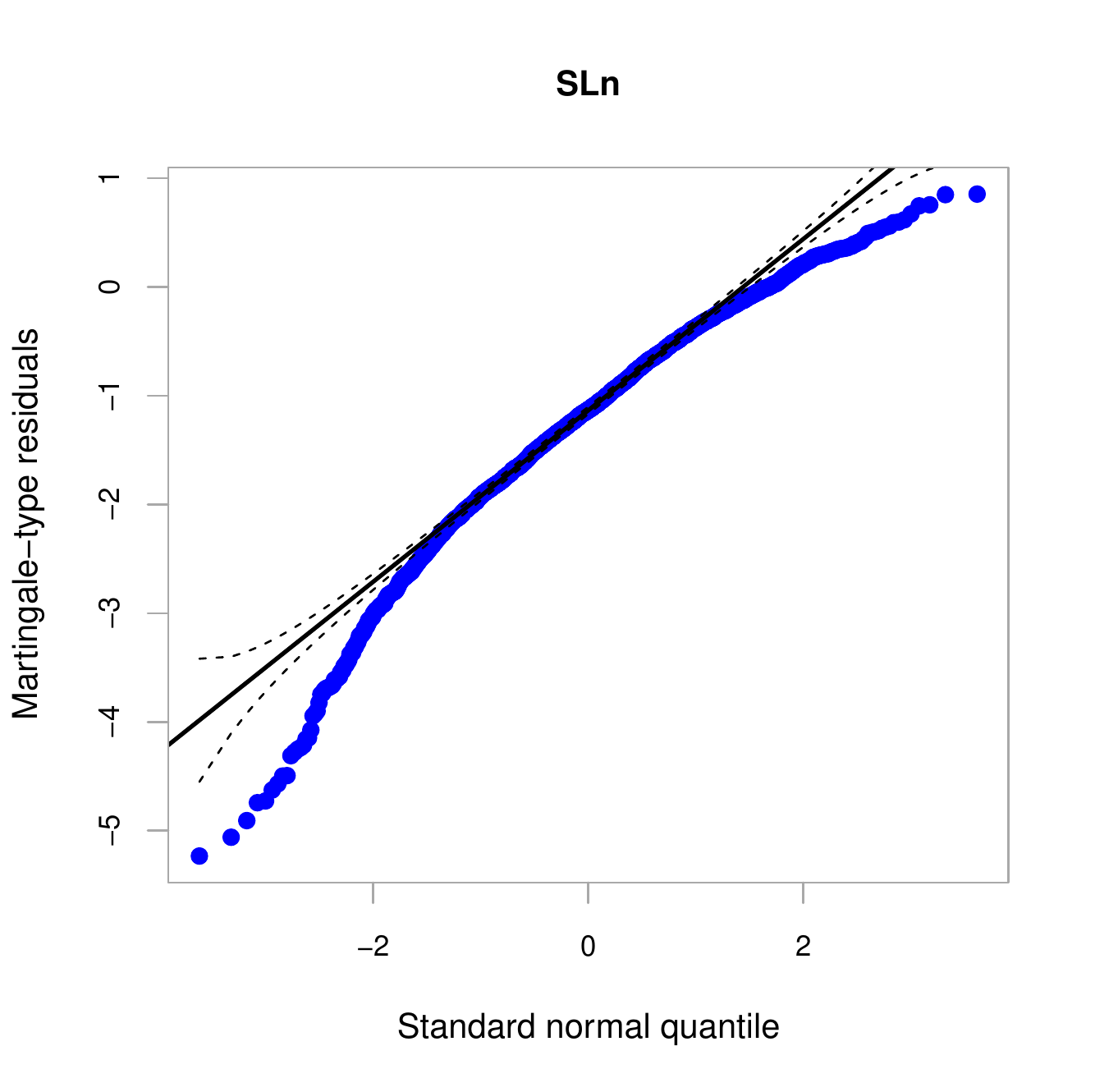} \includegraphics[scale=0.5]{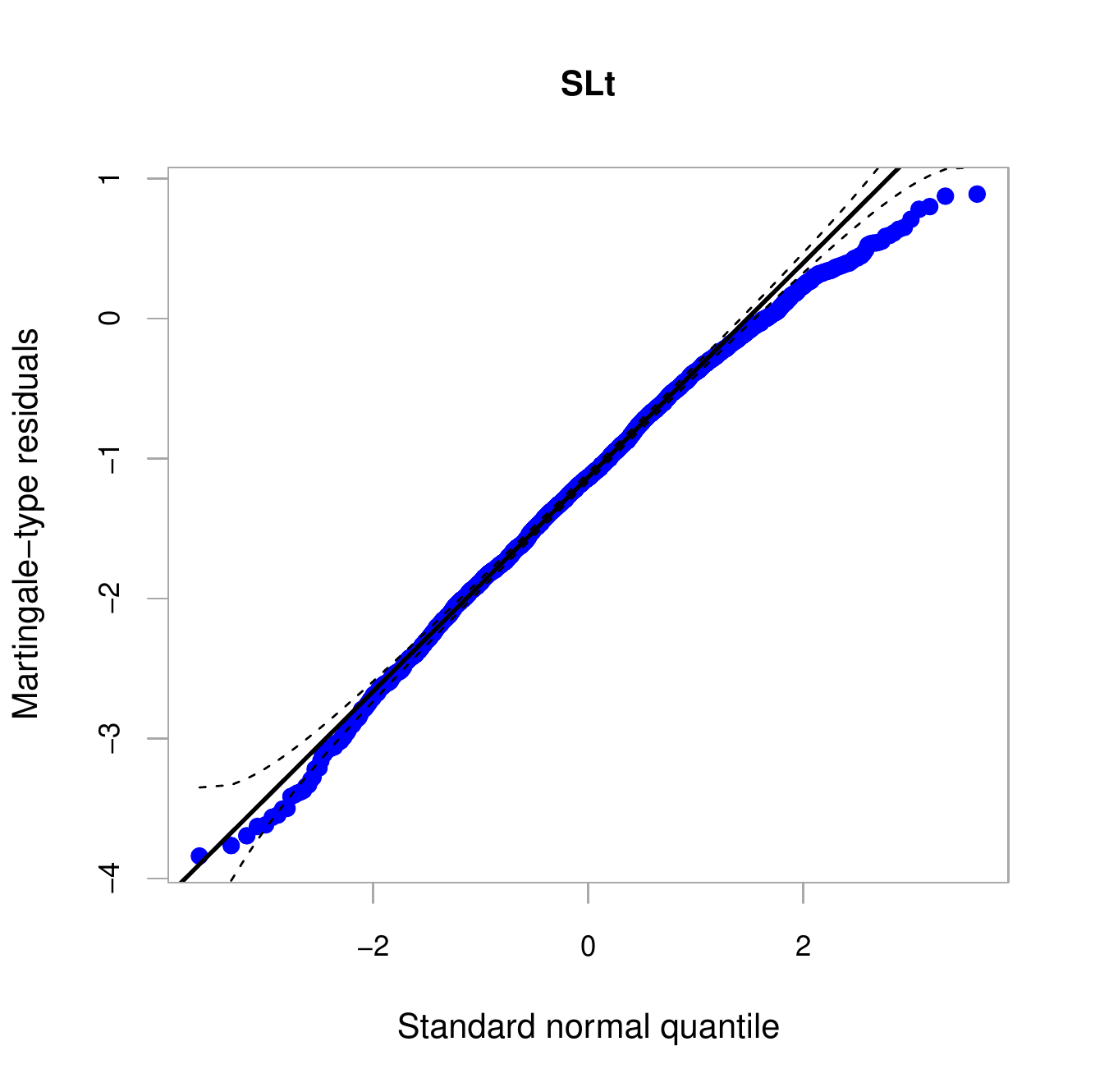}\\
		\caption{Ambulatory expenditures data. Envelopes of the  martingale-type residuals $r_{MT_i}$  for the SLn and SLt models.} \label{fig:envelopes}
	\end{center}
\end{figure}

\begin{table}[!htb]
	\caption{\textbf{Ambulatory expenditures Data}. Parameter estimates and standard errors.}
	\label{tab:expediture}
	\begin{center}
		\begin{tabular}{lcccc}
			\toprule
			& \multicolumn{2}{c}{SLn} & \multicolumn{2}{c}{SLt} \\
			\cmidrule(lr){2-3} \cmidrule(lr){4-5}
			Parameter   & EM    & SE   & EM   & SE \\
			\midrule
			Outcome model (ln expenditures)&\multicolumn{4}{c}{} \\
			\midrule
			$Intercept$ & 5.024 & 0.287 & 5.192 & 0.222	\\
			$age$ & 0.213& 0.024& 0.207 & 0.023 \\
			$female$ &0.353  &0.073 &0.310  &0.060 \\
			$educ$ &0.019  & 0.012  & 0.018& 0.010\\
			$blhisp$ & -0.222&0.065 &-0.195 & 0.059   \\
			$totchr$ & 0.543& 0.054& 0.514 & 0.042 \\
			$ins$ &-0.029 & 0.054&  -0.052& 0.052 \\
			\midrule
			Selection model&\multicolumn{4}{c}{} \\
			\midrule
			$Intercept$ & -0.681 &0.202 & -0.760& 0.217	\\
			$age$ & 0.088&0.027 &   0.099&0.030  \\
			$female$ &0.666& 0.061& 0.732 &  0.067\\
			$educ$ &0.062& 0.013 & 0.065 &0.014  \\
			$blhisp$ &-0.365& 0.063 & -0.396 &  0.067 \\
			$totchr$ & 0.814 &0.069 &0.920&0.084  \\
			$ins$ &0.171& 0.065& 0.182 & 0.070 \\
			$income$ &  0.003&0.001& 0.003 & 0.001 \\
			\midrule
			$\sigma$& 1.270 &0.019& 1.195 &0.023  \\
			$\rho$ & -0.131 & 0.220& -0.321& 0.140\\
			$\nu$ &   &  & 12.928 &  \\
				\midrule
					AIC	& \multicolumn{2}{c}{11674.44} & \multicolumn{2}{c}{11646.15
						
		}\\
				BIC	& \multicolumn{2}{c}{11680.55} & \multicolumn{2}{c}{11652.26

	}\\
			\bottomrule
		\end{tabular}
	\end{center}
\end{table}

\subsection{RAND Health Insurance data}
The second application concerns a study from RAND Health Insurance Experiment (RAND HIE), which is a comprehensive study
of health care cost, utilization, and outcome in the United States. This data set is used by \cite{cameron2009microeconometrics} to analyse how the patient’s use of health services is affected by types of randomly assigned health insurance. More recently, the data were also revisited by \cite{zhao2020new} considering a SLn model.

In our analysis, we choose the same set of variables as \cite{zhao2020new},  we choose the  $\ln$ of the medical expenses of individual
($meddol$) as the outcome variable. The covariates in the outcome equation are:
$$\xp = (1, logc, idp, lpi, fmde, physlm, disea, hlthg, hlthf, hlthp, linc, lfam, educdec, xage, female, child, fchild, black),$$ 
including the log of coinsurance rate plus 1 $(log c=ln (coins+1))$, the dummy
for individual deductible plan ($idp$), the log of participation incentive payment ($lpi$), an artificial variable $fmde$ that is
0 if $idp= 1$ and $ln(max(1,mde/(0.01*coins)))$ otherwise (where mde is the maximum expenditure offer), physical
limitations ($physlm$), the number of chronic diseases ($disea$), dummy variables for good ($hlthg$), fair ($hlthf$), and poor
($hlthp$) self-rated health (where the baseline is excellent self-rated health), the log of family income ($linc$), the log of family size ($lfam$), education of household head in years ($educdec$), age of individual in years ($xage$), a dummy variable for female individuals ($female$), a dummy variable for individuals younger than 18 years ($child$), a dummy variable for female individuals younger than 18 years ($fchild$), and a dummy variable for black household heads (black). The selection
variable  is $binexp$ which indicates whether the medical expenses are positive and without exclusion, we consider that $\bx=\w$.

For our analysis, a subsample was selected so that study year is 2 and educdec is not ‘‘NA’’. Out of 5574 observations, 1293 of $meddol$ (medical expenses) are 0 which means the outcome variable $ln$ of $meddol$ is unobserved, and 4281 of $meddol$ are positive (means that the outcome variable $ln$ of $meddol$ is available). The data is available in the \verb"R" package \verb"sampleSelection".  The results for the SLn model and the SLt model are presented in Table \ref{tab:example2}, as in the previous real analysis, the ML estimates for the SLn model are the same than those reported in \cite{zhao2020new} and also at \url{http://cameron.econ.ucdavis.edu/mmabook/mma16p3selection.txt}. In this case clearly, the 95\% confidence interval of $\rho$ does not contain zero, which suggests the existence of SL effect under both models (SLn and SLt).

The normal probability plot of the $r_{M_{i}}$ residuals with generated envelopes is presented in Figure \ref{fig:envelopes2}. As in the previous example, we observe that the SLt  model
fit the data better than the SLn model, which is corroborated by the relative small value of the degrees of freedom $\nu$, which is $\hat\nu=8.809$.  Moreover,  the AIC and BIC values indicate that the SLt model with heavy tails presents a better fit than the SLn model, due to the departure of the data from normality.

\begin{table}[!ht]
	\caption{\textbf{RAND HIE data.} Parameter estimates and standard errors.}
	\label{tab:example2}
	\begin{center}
		\begin{tabular}{lcccc}
			\toprule
			& \multicolumn{2}{c}{SLn} & \multicolumn{2}{c}{SLt} \\
			\cmidrule(lr){2-3} \cmidrule(lr){4-5}
			Parameter   & EM    & SE   & EM   & SE \\
			\midrule
			Outcome model ($\ln$ of $meddol$)&\multicolumn{4}{c}{} \\
			\midrule
			$Intercept$ & 2.155& 0.253 &2.358  &  0.242 \\
			$logc$ & -0.073& 0.033 & -0.067 & 0.032   \\
			$idp$ &-0.146& 0.063 & -0.152 & 0.061  \\
			$lpi$ & 0.014 & 0.011 &0.014  &0.010   \\
			$fmde$ & -0.024&  0.019&-0.028 &0.018   \\
			$physlm$ &0.350 &0.073  &0.339  & 0.070  \\
			$disea$ & 0.028&  0.004 & 0.028 &  0.0036 \\
			$hlthg$ &0.156 & 0.052 &  0.145 & 0.050  \\
			$hlthf$ &0.442 &  0.092& 0.462 &  0.089 \\
			$hlthp$ &  0.989&  0.167 & 0.881 & 0.165  \\
			$linc$ &0.120& 0.024 & 0.110 &  0.023 \\
			$lfam$ & -0.157 &  0.048 &  -0.180& 0.047  \\
			$educdec$ & 0.017 &  0.009&  0.016 &  0.009 \\
			$xage$ &0.006 &0.002 & 0.005 & 0.002  \\
			$female$ & 0.540& 0.061 & 0.503 & 0.060  \\
			$child$ & -0.202& 0.098 & -0.192 &  0.093 \\
			$fchild$ &-0.554 & 0.100 & -0.526 & 0.095 \\
			$black$&-0.518 &  0.070 &   -0.502& 0.072  \\
			\midrule
			Selection model&\multicolumn{4}{c}{} \\
			\midrule
			$Intercept$ &-0.220 &0.191  &  -0.228&0.207   \\
			$logc$ &-0.108 & 0.025 &  -0.129& 0.028   \\
			$idp$ & -0.110 & 0.048 & -0.105 &0.053   \\
			$lpi$ & 0.030& 0.009 & 0.033 & 0.010  \\
			$fmde$ & 0.002& 0.016 &  0.005 &  0.018 \\
			$physlm$ &0.285 & 0.073 &0.335  & 0.084 \\
			$disea$ & 0.021& 0.004 & 0.024 & 0.004  \\
			$hlthg$ & 0.056&  0.043&  0.055 & 0.047  \\
			$hlthf$ &  0.223& 0.082 & 0.2467&   0.091\\
			$hlthp$ & 0.796& 0.187 & 0.904  &   0.230 \\
			$linc$ & 0.055&0.017  & 0.054 & 0.018  \\
			$lfam$ & -0.032 &0.040  & -0.041 & 0.044  \\
			$educdec$ & 0.032 & 0.008 &0.037 &  0.008  \\
			$xage$ &-0.001 & 0.002 & -0.001 &  0.002 \\
			$female$ & 0.413& 0.053 &  0.463 &  0.059 \\
			$child$ &0.059 &0.079  &0.082  &  0.087 \\
			$fchild$ &-0.401&0.079  & -0.456 & 0.087  \\
			$black$& -0.587& 0.051 & -0.646 & 0.055  \\
			\midrule
			$\sigma$ &1.570 & 0.027&  1.374 &0.027   \\
			$\rho$ & 0.736 & 0.037& 0.667 & 0.047\\
			$\nu$ & & & 8.809 &  \\
			\midrule
				AIC	& \multicolumn{2}{c}{  20342.22} & \multicolumn{2}{c}{20284.12}\\
			BIC	& \multicolumn{2}{c}{20348.85} & \multicolumn{2}{c}{ 20290.75
}\\
			\bottomrule
		\end{tabular}
	\end{center}
\end{table}

\begin{figure}[!ht]
	\begin{center}
		\centering \hspace {1cm}\centering \\
		\includegraphics[scale=0.5]{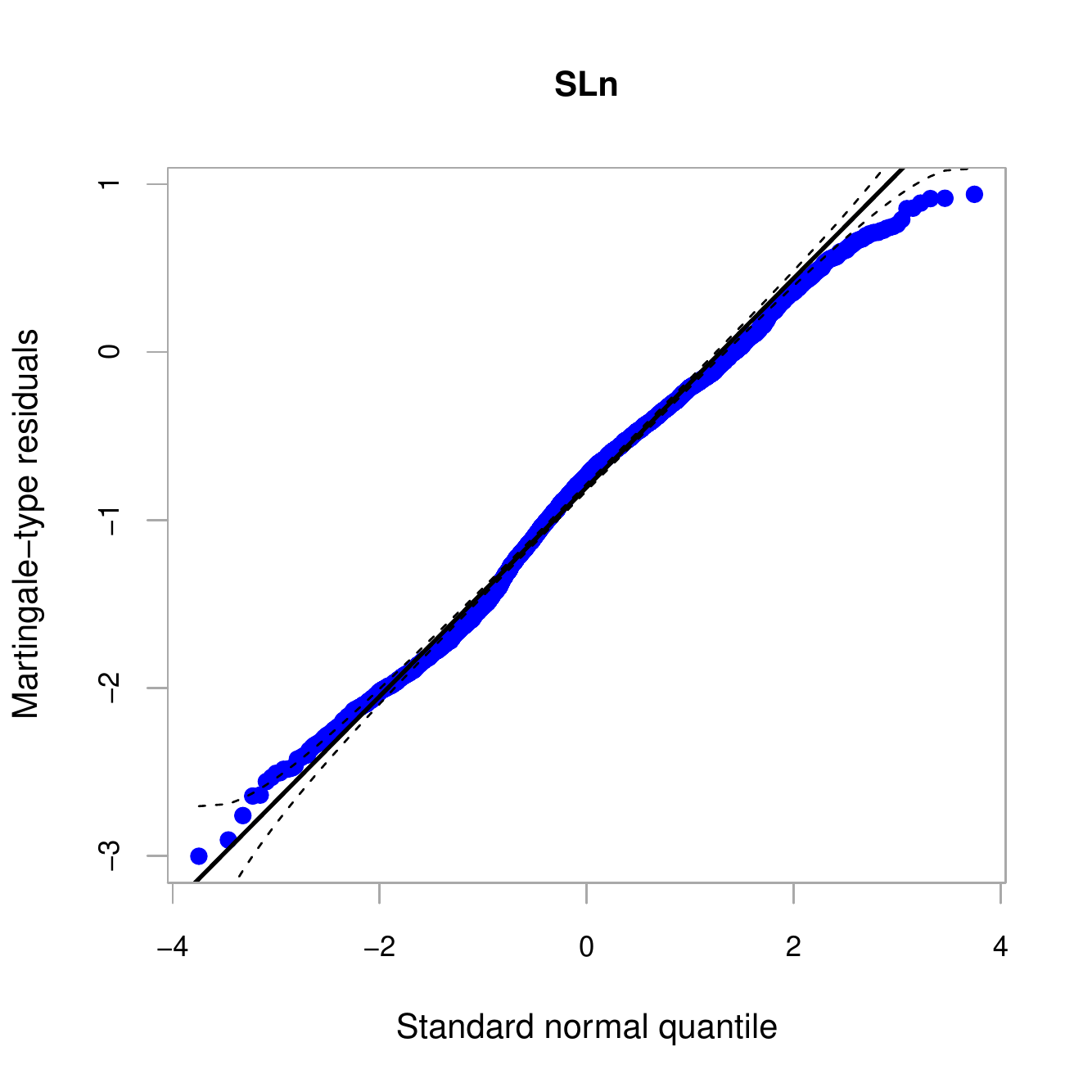} \includegraphics[scale=0.5]{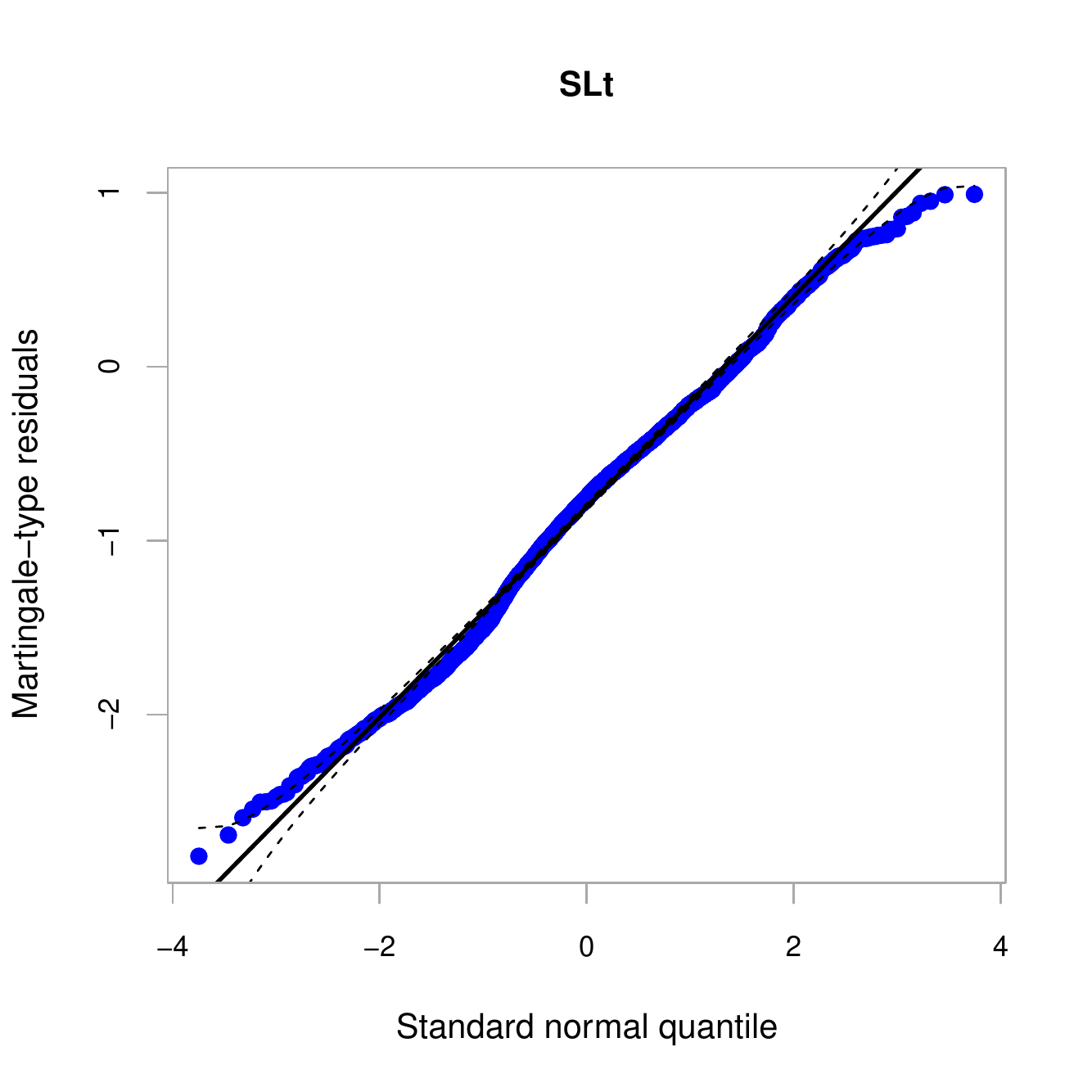}\\
		\caption{RAND HIE data. Envelopes of the  martingale-type residuals $r_{MT_i}$  for the SLn and SLt models.} \label{fig:envelopes2}
	\end{center}
\end{figure}

\section{Conclusions}
\label{sec:6}

In this paper, a novel EM-type algorithm for the SLt model has been developed. In contrast with the existing literature, where to compute the ML estimates, available optimization procedures in standard programs, such as, the \verb"optim" routine in \verb"R" or the \verb"ml" routine in \verb"stata", are used. As discussed by \cite{zhao2020new} (Sec. 6), a disadvantage of direct maximization of the log-likelihood function is that it may not converge unless good starting values are used. Our proposed EM-type algorithm for the SLt model uses closed-form expressions at the E-step, that rely on formulas of the mean and variance of a truncated Student's-$t$ distribution. The general formulas for these moments were derived efficiently by \cite{GalarzaCran}, for which we use the \verb"MomTrunc" package in \verb"R". We also propose a slight modification to the EM-type algorithm proposed by \cite{zhao2020new}, where the parameters in the M-step are updated
by considering the outcome and sample selection as missing data. It is important to point out that the proposed EM algorithm perform more robustly than direct maximization, but is computationally costlier.  The analysis of two real data sets provide strong evidence about the usefulness and effectiveness of our proposal. Moreover, intensive simulation studies show the vulnerability of the SLn model, as well as, the robustness of the SLt model.

A promising avenue for future research is to consider a generalization of the SLt model to the scale mixtures of skew-normal (SMSN) distribution \citep{Lachos_Ghosh_Arellano_2009}, this rich family of SMSN distributions include some well-known multivariate asymmetric heavy-tailed and symmetric distributions, such as, the skew-t \citep{AzzaliniC1999} and the family of scale-mixture of normal distributions \citep{Lange1993}. Another possible extension, includes likelihood-based treatment for the multivariate SL model \citep{tauchmann2010consistency}.
\bigskip

\noindent \textbf{Acknowledgments}\\
This paper was written while Marcos O. Prates was a visiting professor in the Department of Statistics at the University of Connecticut (UConn). In addition to the support of UConn, the professor would like also to thank Conselho Nacional de Desenvolvimento Científico e Tecnológico (CNPq) for partial financial support.


\bibliographystyle{chicago}
\bibliography{bibliornl}

\end{document}